\documentclass[acmsmall,screen]{acmart}
\usepackage{balance}
\usepackage{prftree}
\usepackage{syntax}
\usepackage{stmaryrd}
\usepackage{mathtools}
\usepackage{subcaption}
\usepackage[most]{tcolorbox}

\tcbset{
    boxmath/.style 2 args={%
        enhanced, colback=white,
        colframe=#1, remember as=#2, size=fbox}
    }

\usepackage{anyfontsize}

\title{Explicit Refinement Types}
\author{Jad Elkhaleq Ghalayini}
\affiliation{%
    \institution{University of Cambridge}
    \department{Department of Computer Science and Technology}
    \city{Cambridge}
    \country{United Kingdom}
}
\email{jeg74@cl.cam.ac.uk}
\author{Neel Krishnaswami}
\affiliation{%
    \institution{University of Cambridge}
    \department{Department of Computer Science and Technology}
    \city{Cambridge}
    \country{United Kingdom}
}
\email{nk480@cl.cam.ac.uk}

\setcopyright{rightsretained}

\acmPrice{}
\acmDOI{10.1145/3607837}
\acmYear{2023}
\copyrightyear{2023}
\acmSubmissionID{icfp23main-p19-p}
\acmJournal{PACMPL}
\acmVolume{7}
\acmNumber{ICFP}
\acmArticle{195}
\acmMonth{8}
\received{2023-03-01}
\received[accepted]{2023-06-27}



\newcommand{\mc}[1]{\mathcal{#1}}
\newcommand{\mb}[1]{\mathbf{#1}}
\newcommand{\mbb}[1]{\mathbb{#1}}

\newcommand{\ms}[1]{\ensuremath{\mathsf{#1}}}

\newcommand{\tty}[1]{\texttt{#1}}
\newcommand{\rle}[1]{{\scriptsize\textsf{#1}}}
\newcommand{\brle}[1]{{\small\textsf{#1}}}

\newcommand{\nats}{\mbb{N}}

\newcommand{\ints}{\mbb{Z}}

\newcommand{\lrset}[3]{\left\{{#1}: {#2} \mid {#3}\right\}}
\newcommand{\lrasm}[3]{({#1}: {#2}) \Rightarrow {#3}}
\newcommand{\lrimp}[3]{({#1}: {#2}) \Rightarrow {#3}}
\newcommand{\upg}[1]{{#1}^{\uparrow}}
\newcommand{\dng}[1]{{#1}^{\downarrow}}
\newcommand{\subctx}[2]{{#2} \leq {#1}}
\newcommand{\dnt}[1]{\llbracket{#1}\rrbracket}
\newcommand{\ers}[1]{|{#1}|}

\newcommand{\ednt}[1]{\left\llbracket{#1}\right\rrbracket}
\newcommand{\lrnt}{\vdash}
\newcommand{\stnt}{\vdash_{\lambda}}
\newcommand{\eqop}[1]{\;{=}(#1)\;}
\newcommand{\stlc}{\(\lambda_{\ms{stlc}}\)}
\newcommand{\ert}{{\ensuremath{\lambda_{\ms{ert}}}}}
\newcommand{\tystlc}{\ms{Type}_{\lambda}}
\newcommand{\tmstlc}{\ms{Term}_{\lambda}}

\newcommand{\formalization}[2]{\emph{\small Formalized as: \tty{#1}, \tty{#2}}}

\newcommand\figstrut[2]{
  \dimen0=#1%
  \advance\dimen0 by -#2%
  \divide\dimen0 by -2%
  \dimen1=#1%
  \advance\dimen1 by \dimen0%
  \vrule height \dimen1 depth \dimen0 width 0pt\relax%
}

\newcounter{todos}
\newcommand{\TODO}[1]{{
  \stepcounter{todos}
  \begin{center}\large{\textcolor{red}{\textbf{TODO \arabic{todos}:} #1}}\end{center}
}}

\renewcommand{\TODO}[1]{}

\begin{document}

\begin{abstract}
  We present \ert{}, a type theory supporting refinement types with \emph{explicit proofs}. Instead of solving refinement constraints with an SMT solver like DML and Liquid Haskell, our system requires and permits programmers to embed proofs of properties within the program text, letting us support a rich logic of properties including quantifiers and induction. We show that the type system is sound by showing that every refined program erases to a simply-typed program, and by means of a denotational semantics, we show that every erased program has all of the properties demanded by its refined type. All of our proofs are formalised in Lean 4. 
\end{abstract}

\maketitle

\section{Introduction}

Refinement typing extends an underlying type system with the ability to associate types with logical predicates on their inhabitants, constructing, for example, the type of nonempty lists or integers between three and seven. Most such systems support type dependency between the input and output types of functions, allowing us to give specifications like "the output of this function is the type of integers greater than the input." The core concept underlying type checkers for refinement types is that of logical entailment. If we have a value satisfying a predicate \(P\) and require a value satisfying \(Q\), we require that \(P\) entails \(Q\) for our program to typecheck; this then reduces the typechecking problem to typechecking in our underlying type system plus discharging a set of entailment obligations, called our verification conditions. Given an appropriate choice of logic for our predicates, SMT solvers provide a highly effective way of automatically discharging verification conditions: this allows us to use refinement types without dealing with the bookkeeping details required by manual proofs. Combined with type and annotation inference (in which we also infer refinements), refinement types allow verifying nontrivial properties of complex programs with a minimal annotation burden, making them more appealing for use as part of a practical software development workflow — for example, the Liquid Types implementation of refinement typing for ML required a manual annotation burden of 1\% to prove the DML array benchmarks safe~\cite{liquid-types,dml}.

However, refinement typing's reliance on automation is a double-edged
sword: while it makes refinement types practical for usage in
real-world contexts, it also creates a hard ceiling for
expressiveness, especially if we want to use quantifiers, which would
make the SMT problem undecidable. Rather than carefully massaging
annotations into forms satisfying the particular SMT solver that is in use,
it makes sense to let programmers provide manual proofs for the
(hopefully, few) cases where the solver gets stuck: in utopia,
humans would prove interesting things, and machines would handle the
bookkeeping.

However, it is nontrivial to figure out how to add explicit proofs
into a system, because the semantics of refinement type systems
is subtle. So before we can add the capability to move freely between
explicit proofs and automation, we need to know what manual proofs
should look like! To achieve this goal, we introduce a system of
explicit refinement types, \ert{}, in which all proofs are manual, to
explore the design space. Since our proofs are entirely manual, our
refinement logic can be extremely rich, and, in particular, we support
full first-order logic with quantifiers.

While the presence of explicit proofs means that functions and
propositions look dependently typed, \ert{} is not a traditional
dependent type theory. In particular, we maintain the refinement
discipline that "fancy types" can always be erased, leaving behind a
simply-typed skeleton. Furthermore, \ert{} has no judgemental
equality -- there is no reduction in types. 

\noindent \textbf{Contributions}

\begin{itemize}
    \item We take a simply-typed effectful lambda calculus, \stlc{} (in section~\ref{ssec:stlc}), and add a refinement type discipline to it, to obtain the \ert{} language (in Section~\ref{sec:formalization}).
    \item We support a rich logic of properties, including full first-order quantifiers, as well as ghost variables/arguments (see section~\ref{ssec:typing-rules}). It does not rely on automated proof, and instead features \emph{explicit} proofs of all properties.
    % \item Somewhere: this type system is very rich, but UNLIKE dependent
    %   type systems, does not have or need a judgemental equality. (Eventually cite Zombie and objective type theory)
    \item We show (in section~\ref{ssec:metatheory}) that this type system satisfies the expected properties of a type system, such as the syntactic substitution property.   
    \item In section~\ref{sec:semantics}, we give a denotational semantics for both the simply-typed and refined calculi, and we prove the soundness of the intepretations. Using this, we establish semantic regularity of typing, which shows that every program respects the refinement discipline.
    \item We describe our mechanization of the semantics and proofs in Lean 4 in section~\ref{sec:lean4}. The proofs are available in the supplementary material.
\end{itemize}

\section{Refinement Types}

\subsection{Pragmatics of Refinement Types}

\label{sec:background}

Consider a simply-typed functional programming language. In many cases, the valid inputs for a function are not the entire input type but rather a subset of the input; for example, consider the classic function \tty{head : List A \(\to\) A}. 

Note that we cannot define this as a total function for an arbitrary type \(A\); we must either provide a default value, change the return type, or use some system of exceptions/partial functions. 

Refinement types give us native support for types guaranteeing an invariant about their inhabitants by allowing us to associate a base type with a predicate. For example, we could represent the types of natural numbers and nonzero integers (in a Liquid-Haskell-like notation) as: 
\begin{equation}
  \begin{aligned}
    & \tty{type Nat = \{v: Int | v >= 0\}} \\
    & \tty{type Nonzero = \{v: Int | v <> 0\}}
\end{aligned}
\end{equation}
We could then write the type signature of division as 
\begin{equation}
    \tty{div : Int \(\to\) \{v: Int | v <> 0\} \(\to\) Int}
\end{equation}
By allowing dependency, where the output type of a function is allowed to depend on the value of its arguments, we can use such functionality to encode the specification of operations as well, as in
\begin{equation}
    \tty{eq : x:Int \(\to\) y:Int \(\to\) \{b: Bool | (x = y) = b\}}
\end{equation}
Note that these types all use equality and inequality. This is not the \emph{judgemental} equality of dependent type theory, but rather a \emph{propositional} equality that can occur inside types. We could now attempt to implement a safe-division function on the natural numbers as follows
\begin{equation}
    \begin{array}{l}
        \tty{safeDiv : Nat \(\to\) Nat \(\to\) Nat} \\
        \tty{safeDiv}\;n\;m = \mb{if}\;\tty{eq}\;m\;0\;
            \mb{then}\;0\;
            \mb{else}\;\tty{div}\;n\;m
    \end{array}
    \label{ex:safediv}
\end{equation}
A refinement type system reduces the problem of type-checking to the question of whether a set of logical \emph{verification conditions} holds. 
In this case, all of the propositions are decidable. Unfortunately, with more general specifications, checking these conditions quickly becomes undecidable. For example, if we permitted equations between arbitrary polynomials, then Hilbert's tenth problem -- the solution of Diophantine equatiosn -- can easily be encoded. However, by appropriately restricting the logic of assertions, we can reduce the problem of checking verification conditions to a decidable fragment of logic that, while NP-hard (or worse), can usually be solved very effectively in practice.

In particular, many languages supporting refinement types require refinements to be expressions in the quantifier-free fragment of first-order logic, with atoms restricted to carefully chosen ground theories with efficient decision procedures. In Liquid Haskell, for example, formulas in refinements are restricted to formulas in QF-UFLIA \cite{liquid-haskell}: the \emph{quantifier-free} theory of uninterpreted functions with linear integer arithmetic. While such a system may, at first glance, seem very limited, it is possible to prove very sophisticated properties of real programs with it. In particular, with clever function definitions and termination checking (which generates a separate set of verification conditions to ensure some measure on recursive calls to a function decreases), it is even possible to perform proofs by induction in a mostly-automated fashion \cite{refinement-tut}.

However, it can be challenging to formulate definitions in such a limited fragment of logic, and many properties (such as monotonicity of functions or relational composition) require the use of quantifiers to be stated naturally. Unfortunately, attempts to extend solvers to support more advanced features like quantifiers usually lead to very unreliable performance.

Another issue is that systems designed around off-the-shelf solvers often do not satisfy the \emph{de-Bruijn criterion}: their trusted code-base, rather than being restricted to a small, easily-trusted kernel, instead will often consist of an entire solver and associated tooling! Unfortunately, SMT solvers are incredibly complex pieces of software and hence are very likely to have soundness bugs, as recent research on fuzzing state-of-the-art solvers like CVC4 and Z3 for bugs tends to show \cite{opfuzz}.

In our work, we build a refinement type system, which replaces the use of automated solvers with explicit proofs. This permits (at the price of writing proofs) the free use of quantifiers in propositions, while also having a trusted codebase small enough to be formally verified.

\subsection{Semantics of Refinement Types}
\label{subsec:refsem}
  
To understand what refinement types are semantically, it is worth
recalling that there are two main styles of doing denotational
semantics, which Reynolds dubbed "intrinsic" and "extrinsic"~\citep{reynolds-intrinsic}.

The intrinsic style is the usual style of categorical semantics -- we
find some category where types and contexts are objects, and a term
$\Gamma \vdash e : A$ is interpreted as an element of the hom-set
$\mathrm{Hom}(\Gamma, A)$. (Dependency does make things more
complicated technically, but not in any conceptually essential
fashion.) In the intrinsic style, only well-typed terms have
denotations, and ill-typed terms are grammatically ill-formed and do
not have a semantics at all. Another way of putting this is that 
intrinsic semantics interprets \emph{typing derivations} rather than 
\emph{terms}. 

In the extrinsic style, the interpretation function is defined upon
raw terms, and ill-typed programs \emph{do} get a semantics. Types are
then defined as retracts of the underlying semantic object
interpreting the raw terms. A good example of this style of semantics
is Milner's proof of type safety for ML in \cite{milner-safety}: he
gave a denotational semantics for an untyped lambda calculus, and then
gave an interpretation of types by logical relations over this untyped
calculus, which let him use the fundamental lemma to extract type
safety. Logical relations (and realizabilty methods in general) are
typical examples of extrinsic semantics. Normally type theorists elide
the distinction between these two styles of semantics, because we
prove coherence: that the same term cannot have two derivations with
different semantics. This is called the "bracketing theorem" in
\citet{reynolds-intrinsic} and \citet{ftrs}. 

Refinement types puts these two styles of semantics together in a
different way. We start with a unrefined, base language with an
intrinsic semantics, and then define a \emph{second} system of types
as a family of retracts over the base types. In \ert{}, we use a
monadic lambda-calculus as the base language, and define our system of
refinements over this calculus, with the refinements structured as a
fibration over the base calculus. We deliberately avoided
foregrounding the categorical machinery in our paper to make it more
accessible, though we do direct readers to \citet{ftrs}'s
POPL paper (which deeply influenced our design).

Another, perhaps more familiar, instance of this framework are Hoare
logic and separation logic, which can be understood as refinement type
systems over a base "uni-typed" imperative language. In fact, the
distinction in Hoare logic between "logical variables" (which appear
in specifications) and program variables is (in semantic terms)
precisely the same as the distinction between logical and
computational terms in \ert{} -- which we will discuss further in the
next section. 

\section{Explicit refinement types}

\label{sec:examples}

In this section, we introduce our system of \emph{explicit refinement
  types}, \ert{}, which we construct by enriching the
simply-typed lambda calculus with proofs, intersection types,
and union types. We recover a computational interpretation of
terms in our calculus by recursively erasing the logical
information added, yielding back simply-typed \stlc{} terms. The presence
of proofs allows us to track logical facts,
and by pairing a term with a proof of its property, form subset
types. We also support a general form of intersection and union type,
to allow us to pass around terms used only in proofs in a way that is
guaranteed to be \emph{computationally irrelevant}, which is both a
significant performance concern and allows type signatures to more
clearly express the programmer's intent. For example, consider the
following definition of a vector type:
\begin{equation}
    \ms{Vec}\;A\;n \equiv \{\ell: \ms{List}\;A \mid \ms{List.len}\;\ell = n\}
    \label{eqn:vec-def}
\end{equation}
In particular, we define a vector of length \(n\) as a list paired
with the information that the list has length \(n\). That is, a vector
is a subset type: a base type \(\ms{List}\;A\) paired with a
proposition \(P(\ell)\) (here \(\ms{len}\;\ell = n\)), which we will
interpret as containing all elements \(\ell\) of the base type
satisfying \(P(\ell)\). In general, we write such types as
\(\{x: X \mid P(x)\}\), and introduce them with the form \(\{e, p\}\),
where \(e\) is of type \(X\) and \(p\) is a proof of \(P(e)\).  We
define a length function on vectors as follows:
\begin{equation}
  \begin{array}{lclll}
        \ms{Vec.len} & : & \forall n: \nats, & \ms{Vec}\;A\;n \to & \nats \\
        \ms{Vec.len} & \equiv & \hat{\lambda} \|n: \nats\|, & \lambda v: \ms{Vec}\;A\;n, & 
            \ms{let}\;\{\ell, p\} = v\;\ms{in}\;\ms{List.len}\;\ell
            
    \end{array}
    \label{eqn:vec-len}
\end{equation}
Let us break this definition down, starting from the signature. \(\ms{Vec.len}\) begins by universally quantifying over a natural number \(n\) with the quantifier \(\forall n : \nats\), and then has a function type \(\ms{Vec}\;A\;n \to \nats\). We can interpret this as saying that for every \(n\), \(\ms{Vec.len}\) takes vectors of length \(n\) to a natural number.

The $\forall$ quantifier in the type of $\ms{Vec.len}$ behaves like the quantifiers in ML-style polymorphism, rather than the pi-type of dependent type theory. It indicates that the \emph{same} function can handle vectors of any length, with no explicit branching on the length. In contrast, a pi-type can lets the function body compute with the natural number argument. 

An explicit binding, written \(\lambda\|n: \nats\|\), represents this
quantifier in the actual definition; we call a variable binding
surrounded by double bars (e.g., \(\|x: A\|\)) a \emph{ghost
  binding}. Moving inwards, we have a function type with input
\(\ms{Vec}\;A\;n\) and output \(\nats\); this corresponds to the
lambda-expression "\(\lambda v: \ms{Vec}\;A\;n, E\)" in the
definition, where
\(E \equiv \ms{let}\;\{\ell, \_\} = v\;\ms{in}\;\ms{len}\;\ell\) is an
expression of type \(\nats\).  Breaking down \(E\), we must first
explicitly destructure our vector \(v\) into its components, a list
\(\ell\) and a proof, \(p\) that \(\ell\) is of length \(n\). Unlike
in refinement type systems like DML and Liquid Haskell, this is
explicit, with no entailment-based subtyping.

The definition in equation~\ref{eqn:vec-len} ignores the proof component of the let-binding and hence the refinement information carried by our type system. One way to use the proof information would be to have a definition for \(\ms{Vec.len}\), which promises to return an integer equal to the length. 
\begin{equation}
  \begin{array}{lclll}
    \ms{Vec.len'} & : & \forall n: \nats,
                      & (v: \ms{Vec}\;A\;n) \to 
                      & \{x: \nats \mid x = n\} \\
    \ms{Vec.len'} & \equiv{} & 
        \hat{\lambda}\|n: \nats\|, & \lambda v: \ms{Vec}\;A\;n, & 
            \ms{let}\;\{\ell, p\}=v\;\ms{in}\;\{\ms{List.len}\;\ell, p\}
            
    \end{array}
    \label{eqn:vec-len2}
\end{equation}
However, one helpful feature of \ert{} is that even in this case, we can prove facts about our definitions by writing freestanding proofs about them rather than cramming every possible fact we could want into our type signature. For example, we could give a proof a proposition that the definition in equation~\ref{eqn:vec-len} is correct as follows:
\begin{equation}
  \begin{array}{ll}
    \ms{Vec.len\_def} : &    \forall n: \nats, \forall v: \ms{Vec}\;A\;n, 
                               \ms{Vec.len}\;\|n\|\;v = n \\
    \ms{Vec.len\_def} \equiv{} 
        & \hat{\lambda} \|n: \nats\|, 
        \hat{\lambda} \|v: \ms{Vec}\;A\;n\|,
        \ms{let}\;\{\ell, p\} = v\;\ms{in}\;
        \\&\ms{trans}[ \ms{Vec.len}\;\|n\|\;\{\ell, p\} 
        \\&\qquad \eqop{\beta_{\ms{ir}}} 
            (\lambda v, 
                \ms{let}\;\{\ell, \_\} = v\;\ms{in}\;
                    \ms{List.len}\;\ell)\{\ell, p\}
        \\&\qquad \eqop{\beta_{\ms{ty}}} 
                (\ms{let}\;\{\ell, \_\} = \{\ell, p\}\;\ms{in}\;
                    \ms{List.len}\;\ell)
        \\&\qquad \eqop{\beta_{\ms{set}}}\;\ms{List.len}\;\ell 
        \\&\qquad \eqop{p}\;n] 
    \end{array}
\end{equation}
Let us break this definition down, again starting with the
signature. We quantify over both the length \(n : \nats\) and the
vector \(v : \ms{Vec}\;A\;n\), and then assert the equality
proposition \(\ms{Vec.len}\;\|n\|\;v = n\). In the proofs, both
universal quantifiers are introduced by ghost lambdas
"\(\hat{\lambda}\|n: \nats\|\)" and
"\(\hat{\lambda}\|v: \ms{Vec}\;A\;n\|\)". However, we use
\(\hat{\lambda}\) instead of \(\lambda\) because we are proving a
universally quantified \emph{proposition} rather than forming a
\emph{term} of intersection type. The proof of equality is a term of
the form \(\ms{trans}[...]\), which represents an Agda-style syntax
sugar for equational reasoning. In particular, if
\(\ms{trans}_{p, q, r}: p = q \to q = r \to p = r\) is the
transitivity rule, then we have the desugaring
\begin{equation}
    \ms{trans}[
        x_0 
        \eqop{p_0} x_1
        \eqop{p_1} x_2
        \;...\; 
        x_n \eqop{p_n} x_{n + 1}
    ] \equiv
    \ms{trans}_{x_0, x_1, x_n}\;p_0 (\ms{trans}_{x_1, x_2, x_n}\;p_1\;(...))
\end{equation}
where each \(p_i\) is evidence of the proposition \(x_i = x_{i + 1}\).

Examining the proof of equation~\ref{eqn:vec-def}, we see that every piece of evidence used except one is of the form "\(\beta_{\ms{something}}\)". These are explicit \(\beta\)-reduction proofs, which are necessary since our calculus does not include a notion of judgemental equality: type equality is simply \(\alpha\)-equivalence.
\footnote{
    The actual \(\beta\)-reduction rules in our calculus, e.g. \hyperref[fig:ert-axiom-rule]{\(\beta_{\ms{ty}}\)}, require explicit annotations to be unambiguous. We omit these here for space and clarity. 
}
While this dramatically simplifies the meta-theory, this can make even simple proofs very long. 
%since we need to write out all our reductions as applications of axioms
In our examples, we implement the pattern of repeatedly applying a \(\beta\)-reduction 
%(which, essentially, is what \(\beta\)-conversion is) 
as syntactic sugar. Writing it as \(\beta\) (pronounced "by beta"), we get the much simpler proof 
\begin{equation}
  \begin{array}{ll}
    \ms{Vec.len\_def} & :\forall n: \nats, \forall v: \ms{Vec}\;A\;n, \ms{len}\;v = n \\
        \ms{Vec.len\_def} & \equiv \hat{\lambda} \|n: \nats\|, \hat{\lambda} \|v: \ms{Vec}\;A\;n\|,
        \\& \hspace{0.5cm} \ms{let}\;\{\ell, p\} = v\;\ms{in}\;
            \ms{trans}[\ms{len}\;n\;\{\ell, p\} 
            \eqop{\beta} \ms{len}\;\ell 
            \eqop{p} n]
    \end{array}
\end{equation}
% Through more complex tactics, it is possible to obtain more complex automatic equality checking procedures specialized to the task at hand (for example, taking into account associativity/commutativity axioms).

However, these definitions raise a question: why not simply write 
\begin{equation}
    \ms{Vec.len} \equiv \lambda \|n: \nats\|,\lambda v: \ms{Vec}\;A\;n, n
    \label{eqn:bad-vec-len}
\end{equation}
The problem with the above definition is that \(n\) is a ghost variable, indicating we may use it in propositions and proofs but not generally in terms.
This is a critical distinction to be able to make: the specification of a program may involve values which we do not want to manipulate at runtime. For example, the correctness proof of a sorting routine might use an inductive datatype of permutations, elements of which could potentially be much larger than the list itself. By making a distinction between ghost and computational variables, we can define an efficient erasure of refined terms (which contain proofs) to simply-typed terms (which do not). For example, we can erase the signature in equation~\ref{eqn:vec-len2} into a simple type:
\begin{align}
    &|\forall n: \nats, (v: \ms{Vec}\;A\;n) \to \{x: \nats \mid x = n\}|
    \\&=
    \mb{1} \to |\{\ell: \ms{List}\;A \mid \ms{len}\;\ell = n\} \to \{x: \nats \mid x = n\}| 
    \tag{unfold, erase quantified variable \(n\)}
    \\&= 
    \mb{1} \to |\{\ell: \ms{List}\;A \mid \ms{len}\;\ell = n\}| \to |\{x: \nats \mid x = n\}| 
    \tag{erasure distributes over function types}
    \\&=
    \mb{1} \to |\ms{List}\;A| \to |\nats|
    \tag{erase subset types to erased base types}
    \\&=
    \mb{1} \to \ms{List}\;|A| \to \nats
    \tag{list erases to list of erased type, \(\nats\) erases to \(\nats\)}
\end{align}
We can then erase the definition as follows: 
\begin{align}
    &|\lambda \|n: \nats\|,\lambda v: \ms{Vec}\;A\;n,
    \ms{let}\;\{\ell, \_\} = v\;\ms{in}\;\ms{len}\;\ell|
    \\ &= \lambda n: \mb{1}, |\lambda v: \{\ell: \ms{List}\;A \mid \ms{len}\;\ell = n\},
        \ms{let}\;\{\ell, \_\} = v\;\ms{in}\;\ms{len}\;\ell|
    \tag{unfold, erase quantified variable \(n\)}
    \\ &= \lambda n: \mb{1}, \lambda v: |\{\ell: \ms{List}\;A \mid \ms{len}\;\ell = n\}|, 
        |\ms{let}\;\{\ell, \_\} = v\;\ms{in}\;\ms{len}\;\ell|
    \tag{erasure distributes over binders}
    \\ &= \lambda n: \mb{1}, \lambda v: |\ms{List}\;A|, 
        |\ms{let}\;\{\ell, \_\} = v\;\ms{in}\;\ms{len}\;\ell|
    \tag{erase subset types to erased base types}
    \\ &= \lambda n: \mb{1}, \lambda v: \ms{List}\;|A|, 
        |\ms{let}\;\{\ell, \_\} = v\;\ms{in}\;\ms{len}\;\ell|
    \tag{list erases to list of erased type}
    \\ &= \lambda n: \mb{1}, \lambda v: \ms{List}\;|A|, 
        \ms{let}\;\ell = |v|\;\ms{in}\;|\ms{len}\;\ell|
    \tag{erase distributes over \(\ms{let}\)}
    \\ &= \lambda n: \mb{1}, \lambda v: \ms{List}\;|A|, 
        \ms{let}\;\ell = v\;\ms{in}\;|\ms{len}\;\ell|
    \tag{variables erase to themselves}
    \\ &= \lambda n: \mb{1}, \lambda v: \ms{List}\;|A|, 
        \ms{let}\;\ell = v\;\ms{in}\;|\ms{len}|\;|\ell|
    \tag{erasure distributes over application}
    \\ &= \lambda n: \mb{1}, \lambda v: \ms{List}\;|A|, 
        \ms{let}\;\ell = v\;\ms{in}\;\ms{len}\;\ell:
        \mb{1} \to \ms{List}\;|A| \to \nats
    \tag{variables, \(\ms{len}\) erase to themselves}
\end{align}
At the type level, we erase any dependency information, leaving us
with simple types. Propositions are all either wholly erased or erased
to the unit type. At the term level, erasure essentially consists of
recursively erasing ghost variables and proofs into units and (in the
case of proofs of falsehood) error stops, yielding a well-typed term
in the simply-typed lambda calculus extended with an error stop effect
(i.e., the exception monad). We will prove in section~\ref{sec:semantics}
that the produced terms are always well-typed.

As expected, we see that erasure sends base types like \(\mb{1}\) and \(\nats\), as well as their literals like \(()\) or \(0\), to themselves. Erasure distributes over (dependent) function types, i.e., \(|(a: A) \to B(a)| = |A| \to |B(a)|\),
and correspondingly is recursively applied to the argument type and result of a lambda function as follows: 
\(|\lambda a: A, b| = \lambda a: |A|, |b|\).
Subset types are erased to the erasure of their base type, i.e., we have
\(|\{x: X \mid P(x)\}| = |X|\).
Erasure of the formation and elimination rules for subset types is likewise as expected, with 
\(|\{x, p\}| = |x|\) 
and \(|\ms{let}\;\{x, p\}=e\;\ms{in}\;e'| = \ms{let}\;x=|e|\;\ms{in}\;|e'|\).
Similarly, universal quantifiers are erased to the unit type \(\mb{1}\) as follows: \(
    |\forall a: A, B(a)| = \mb{1} \to |B(a)|
    \label{eqn:prop-erasure-ex}
    \).
Correspondingly, we erase the introduction and elimination rules for intersection types like so: 
\(|\lambda\|a: A\|, b| = \lambda\_: \mb{1}, |b|\), 
\(|f\;\|a\|| = |f|\;()\). In general, propositions and ghosts in value types (such as subsets) are erased completely, whereas propositions and ghosts in function-style types (such as intersection types) are erased to units to avoid problems with the eager evaluation order of \stlc{}. (See section~\ref{par:units-in-the-erasure} for a more detailed discussion.)

Returning to the original question, erasing the definition in equation~\ref{eqn:bad-vec-len} yields the term \(\lambda n: \mb{1}, \lambda v: \ms{List}\;|A|, n\),
which is not well-typed at \(\mb{1} \to \ms{List}\;|A| \to \nats\).
Another way of thinking about this is that if \(\ms{Vec.len}\) took the length \(n\) as a computational argument, we would never need to call the function -- we would have to have the length in hand to call \(\ms{Vec.len}\) in the first place. However, in our setting, vectors are merely refined lists, which erase into raw lists. A raw list does not carry its length \(n\); but \(n\) is a well-defined property of its specification. Since the specification value \(n\) gets erased to a unit, then a program that wants to compute the length must traverse the list -- i.e., in equation~\ref{eqn:vec-len} we call \(\ms{List.len}\). 

%The most effective algorithm might also be different at different points; for example, given a function \(\ms{cons}: A \to \ms{Vec}\;A\;n \to \ms{Vec}\;A\;(n + 1)\), a vector \(v: \ms{Vec}\;A\;n\), a computational variable $x : \nats$ and a proof \(p: x = n\), we may compute the length of \(\ms{cons}\;a\;v\) by simply calculating \(x + 1\). We then obtain proof of correctness simply by congruence; in real life, what is essentially this optimization can change an \(\mc{O}(n^2)\) program (which must repeatedly re-compute the list length) to \(\mc{O}(n)\).

Another advantage of having explicit proofs as part of a refinement type system is the ability to reuse previous theorems and perform proofs by induction. For a simple example of this, consider the following definition of addition for natural numbers:
\begin{equation}
    n + m \equiv (\ms{natrec}\;n\;(\lambda x,x)\;(\|\ms{succ}\;\_\|, f \mapsto \lambda x,\ms{succ}\;(f\;x)))\;m: \nats
\end{equation}
\(\ms{natrec}\) is the eliminator for the natural numbers, which is defined essentially by iteration, with typing rule \hyperref[fig:ert-term-rule]{\brle{Natrec}}. The eliminator has the standard behavior, given by reduction rules \hyperref[fig:ert-axiom-rule]{\(\beta_{\ms{zero}}\)} and \hyperref[fig:ert-axiom-rule]{\(\beta_{\ms{succ}}\)}, which essentially amount to substituting \(z\) into \(s\) recursively \(n\) times. We can then use these axioms to prove that zero is a left-identity by \(\beta\)-reduction, simply writing \(\ms{zero_{left}} : (\forall n: \nats, 0 + n = n) \equiv \hat{\lambda}\|n: \nats\|, \beta\).
In contrast, we need induction to prove that zero is a right-identity. To perform induction, we introduce the \(\ms{ind}\) eliminator for natural numbers, essentially the propositional version of \(\ms{natrec}\), with typing rule \hyperref[fig:ert-proof-rule]{\brle{Ind}}. We may then write 
\begin{flalign}
    &\begin{aligned}
        &\ms{zero_{right}} : (\forall n: \nats, n + 0 = n) \equiv 
        \\&
        \hspace{1cm}
        \hat{\lambda}\|n: \nats\|,
        \ms{ind} [x \mapsto x + 0 = x]\;n
        \;\beta
        \\& \hspace{2.54cm} (\ms{succ}\;n, u \mapsto
        \ms{trans}[
                        (\ms{succ}\;n) + 0
                        \eqop{\beta} \ms{succ}\;(n + 0)
                        \eqop{u} \ms{succ}\;n
                ]
            )
        ) & 
    \end{aligned}&
\end{flalign}
Induction lets us prove many arithmetic facts without baking them into the type system. For example, we can prove that addition is commutative as follows (see figure~\ref{fig:comm-aux} for helpers), where \(\ms{symm}\) and \(\ms{congr}\) are proofs that equality is symmetric and transitive respectively: 
\begin{flalign}
    &\begin{aligned}
        &\ms{add_{comm}} : (\forall n, m: \nats, n + m = m + n)
        \equiv \hat{\lambda}\|n: \nats\|,
        \\& \quad \ms{ind} [x \mapsto \forall m: \nats, x + m = m + x]\;n
        \;\ms{zero_{comm}}
        \\& \qquad (\ms{succ}\;n, u \mapsto 
        \ms{trans}[\ms{succ}\;n + m
        \\& \qquad\quad       \eqop{\beta} \ms{succ}\;(n + m)
        \\& \qquad\quad       \eqop{\ms{congr}\; u} \ms{succ}\;(m + n)
        \\& \qquad\quad       \eqop{\beta} \ms{succ}\;m + n
        \\& \qquad\quad       \eqop{\ms{symm}\;(\ms{succ_{comm}\;\|m\|\;\|n\|})} m + \ms{succ}\;n
        ]
        ) & 
    \end{aligned}&
    \label{eqn:add-comm}
\end{flalign} 
\begin{figure}
    \begin{align*}
        &\ms{zero_{comm}} : (\forall n: \nats, 0 + n = n + 0)
          \equiv \hat{\lambda}\|n: \nats\|,
            \ms{trans}[0 + n 
                \eqop{\beta} n
                \eqop{\ms{zero_{right}}\;\|n\|} n + 0 ] 
        \\
        &\ms{succ_{right}} : (\forall n, m: \nats, n + \ms{succ}\;m = \ms{succ}\;(n + m))
        \equiv 
        \\& \qquad \hat{\lambda}\|n, m: \nats\|, \ms{ind} [x \mapsto x + (\ms{succ}\;m) = \ms{succ}\;(x + m)]\;n
        \;\beta
        \\& \hspace{2cm}\qquad(\ms{succ}\;n, u \mapsto 
        \ms{trans}[\ms{succ}\;n + \ms{succ}\;m
        \\& \hspace{2cm}\qquad\quad    \eqop{\beta} \ms{succ}\;(n + \ms{succ}\;m)
        \\& \hspace{2cm}\qquad\quad    \eqop{u} \ms{succ}\;(\ms{succ}\;(n + m))
        \\& \hspace{2cm}\qquad\quad    \eqop{\beta} \ms{succ}\;(\ms{succ}\;n + m)
        ])
        )
        \\
        &\ms{succ_{comm}} : (\forall n: \nats, n + \ms{succ}\;m = \ms{succ}\;n + m)
          \equiv \\
          & \qquad \hat{\lambda}\|n: \nats\|,
            \ms{trans}[
                    n + \ms{succ}\;m 
                    \eqop{\ms{succ_{right}}\;\|n\|\;\|m\|} \ms{succ}(n + m) 
                    \eqop{\beta} \ms{succ}\;n + m]
    \end{align*}
    \caption{Helper definitions for equation~\ref{eqn:add-comm}}
    \Description{Helper definitions for equation~\ref{eqn:add-comm}}
    \label{fig:comm-aux}
\end{figure}
Note the ability to reuse theorems we have proved previously. Another
advantage of explicit proofs is that we do not need to encode even
fundamental facts into our core calculus since we can prove them from
a small core of base axioms. Minimizing the number of axioms
simplifies the implementation of the type-checker and reduces the size
of the trusted codebase while allowing the programmer to effectively
write refinements and proofs using facts that the language designer
may not have considered.

\section{Formalization}

\label{sec:formalization}

We give \ert{}'s grammar and typing rules in sections \ref{ssec:grammar} and \ref{ssec:typing-rules}. We then give proofs of some expected metatheoretic properties, such as substitution and regularity, in section~\ref{ssec:metatheory}.

\subsection{Grammar}
\label{ssec:grammar}

The grammar for \ert{} (given in figure~\ref{fig:ert-grammar} in the appendix) consists of four separate syntactic categories: types \(A\), propositions \(\varphi\), terms \(a\), and proofs \(p\). We denote the set of (syntactically) well-formed terms living in each by \(\ms{Type}\), \(\ms{Prop}\), \(\ms{Term}\), and \(\ms{Proof}\), respectively.
We may then define a typing context \(\Gamma\) as a list of computational variables \(x: A\), ghost variables \(\|x: A\|\), and propositional variables \(u: \varphi\), as in figure~\ref{fig:ert-ctx-wf}.
% \begin{figure}
%     \begin{center}    
%         \begin{grammar}
%             <\(\Gamma, \Delta\)> ::= \(\cdot\) 
%             \;|\; <\(\Gamma\)>, <H>

%             <H> ::= <x>: <A>
%             \;|\; \(\|x: A\|\)
%             \;|\; \(p: \varphi\)
%         \end{grammar}
%     \end{center}
%     \caption{Grammar for \ert{} contexts and hypotheses}
%     \Description{Grammar for \ert{} contexts and hypotheses}
%     \label{fig:ctx-grammar}
% \end{figure}
Ghost variables only appear within proofs, types, and propositions, whereas computational variables can occur anywhere, including both computational and logical terms. Typing contexts are telescopic, so types and propositions appearing later in the context may depend on previously defined variables.
We use these syntactic  categories to state the shape of \ert{}'s typing judgments in figure~\ref{fig:ert-struct}.
\begin{figure}
    \begin{center}
        \begingroup
        \renewcommand{\arraystretch}{1.5}
        \begin{tabular}{lr}
            \multicolumn{1}{c}{Judgment} & \multicolumn{1}{c}{Meaning} \\ \hline
            \(\Gamma\;\ms{ok}\) & 
            \(\Gamma\) is a well-formed context \\
            \(\Gamma \lrnt A\;\ms{ty}\) & \(A\) is a well-formed type in \(\Gamma\) \\
            \(\Gamma \lrnt \varphi\;\ms{pr}\) & 
            \(\varphi\) is a well-formed proposition in \(\Gamma\) \\
            \(\Gamma \lrnt a: A\) & 
            \(a\) may consistently be assigned type \(A\) in \(\Gamma\) \\
            \(\Gamma \lrnt p: \varphi\) & 
            \(p\) is a proof of \(\varphi\) in \(\Gamma\)
        \end{tabular}
        \endgroup
    \end{center}
    \caption{\ert{} typing judgements}
    \Description{\ert{} typing judgements}
    \label{fig:ert-struct}
\end{figure}
\noindent
A context is \emph{well-formed} if the types of each of its variables are well-formed in the context made up of all previously defined variables, with the context well-formedness rules in figure~\ref{fig:ert-ctx-wf}.
\begin{figure}
    \begin{equation*}
        \prftree{\cdot\;\ms{ok}} \qquad
        \prftree
            {\Gamma\;\ms{ok}}
            {\Gamma \lrnt A\;\ms{ty}}
            {\Gamma, x: A\;\ms{ok}} \qquad 
        \prftree
            {\Gamma\;\ms{ok}}
            {\Gamma \lrnt A\;\ms{ty}}
            {\Gamma, \|x: A\|\;\ms{ok}} \qquad
        \prftree
            {\Gamma\;\ms{ok}}
            {\Gamma \lrnt \varphi\;\ms{pr}}
            {\Gamma, u: \varphi\;\ms{ok}}
    \end{equation*}
    \caption{\ert{} context well-formedness rules}
    \Description{ERT context well-formedness rules}
    \label{fig:ert-ctx-wf}
\end{figure}
\noindent
The presence of both computational and ghost variables means that our contexts have additional structural properties beyond the usual ones such as weakening and exchange. Since a computational variable can be used in more places than a ghost variable, we introduce the concept of an \emph{upgrade}. When we upgrade a context, some of the ghost variables can be replaced by computational variables with the same name and type. So, for example, the context \(\|x:A\|, y:B\) can be upgraded to \(x:A, y:B\). In figure~\ref{fig:ert-ctx-upgrade}, we formalise this with the rules of the judgment
\(\subctx{\Gamma}{\Delta}\), which reads "\(\Delta\) upgrades \(\Gamma\)".

\begin{figure}[H]
    \begin{equation*}
        \prftree
            {\subctx{\Gamma}{\Delta}}
            {\subctx{\Gamma, \|x: A\|}{\Delta, x: A}} \qquad
        \prftree
            {\subctx{\Gamma}{\Delta}}
            {\subctx{\Gamma, H}{\Delta, H}} \qquad
        \prftree{}
            {\subctx{\cdot}{\cdot}}
    \end{equation*}
    \caption{\ert{} context upgrade rules, where \(H\) ranges over hypotheses \(x: A\), \(\|x: A\|\), \(p: \varphi\)}
    \Description{ERT context upgrade rules, where H ranges over hypotheses}
    \label{fig:ert-ctx-upgrade}
\end{figure}
\noindent
We now define the \emph{upgrade} of \(\Gamma\), written \(\upg{\Gamma}\), to be the context with all ghost variables in \(\Gamma\) replaced by term variables, that is,
\begin{equation}
    \upg{\cdot} = \cdot \qquad
    \upg{(\Gamma, \|x: A\|)} = \upg{\Gamma}, x: A, \qquad
    \upg{(\Gamma, x: A)} = \upg{\Gamma}, x: A, \qquad
    \upg{(\Gamma, u: \varphi)} = \upg{\Gamma}, u: \varphi
\end{equation} 
Note that \(\subctx{\Gamma}{\upg{\Gamma}}\) (but \(\subctx{\upg{\Gamma}}{\Gamma}\) iff \(\Gamma\) contains no ghost variables). A context \(\Delta\) which upgrades \(\Gamma\) types more terms than \(\Gamma\), since we may use a computational variable anywhere a ghost variable is expected, but not vice versa. In particular, we may prove the following lemma: 
\begin{lemma}[Upgrade]
    Given contexts \(\subctx{\Gamma}{\Delta}\), 
    \begin{itemize}
        \item If \(\Gamma \lrnt \varphi\;\ms{pr}\), then \(\Delta \lrnt \varphi\;\ms{pr}\). 
        In particular, if \(\Gamma \lrnt \varphi\;\ms{pr}\), then \(\upg{\Gamma} \lrnt \varphi\;\ms{pr}\). 
        \item If \(\Gamma \lrnt A\;\ms{ty}\), then \(\Delta \lrnt A\;\ms{ty}\).
        In particular, if \(\Gamma \lrnt \varphi\;\ms{ty}\), then \(\upg{\Gamma} \lrnt \varphi\;\ms{ty}\).
        \item If \(\Gamma \lrnt p: \varphi\), then \(\Delta \lrnt p: \varphi\).
        In particular, if \(\Gamma \lrnt p: \varphi\), then \(\upg{\Gamma} \lrnt p: \varphi\).
        \item If \(\Gamma \lrnt a: A\), then \(\Delta \lrnt a: A\).
        In particular, if \(\Gamma \lrnt a: A\), then \(\upg{\Gamma} \lrnt a: A\).
        \item If \(\Gamma\;\ms{ok}\), then \(\Delta\;\ms{ok}\). In particular, if \(\Gamma\;\ms{ok}\), then \(\upg{\Gamma}\;\ms{ok}\).
    \end{itemize}
    \label{lem:upgrade}
\end{lemma}
Since the only difference between computational and ghost variables is that ghosts can't be used in computational terms, this distinction does not matter for proofs, or for proposition- and type-well-formedness. Formally:

\begin{lemma}[Downgrade]
    Given contexts \(\subctx{\Gamma}{\Delta}\), 
    \begin{itemize}
        \item If \(\Delta \lrnt \varphi\;\ms{pr}\), then \(\Gamma \lrnt \varphi\;\ms{pr}\). 
        In particular, if \(\upg{\Gamma} \lrnt \varphi\;\ms{pr}\), then \(\Gamma \lrnt \varphi\;\ms{pr}\). 
        \item If \(\Delta \lrnt A\;\ms{ty}\), then \(\Gamma \lrnt A\;\ms{ty}\).
        In particular, if \(\upg{\Gamma} \lrnt \varphi\;\ms{ty}\), then \(\Gamma \lrnt \varphi\;\ms{ty}\).
        \item If \(\Delta \lrnt p: \varphi\), then \(\Gamma \lrnt p: \varphi\).
        In particular, if \(\upg{\Gamma} \lrnt p: \varphi\), then \(\Gamma \lrnt p: \varphi\).
        \item If \(\Delta\;\ms{ok}\), then \(\Gamma\;\ms{ok}\). In particular, if \(\upg{\Gamma}\;\ms{ok}\), then \(\Gamma\;\ms{ok}\).
    \end{itemize}
    \label{lem:downgrade}
\end{lemma}

\subsection{Typing Rules}

\label{ssec:typing-rules}

The type formation rules for \ert{} are collected in figure~\ref{fig:ert-type-rule}, the term formation rules in figure~\ref{fig:ert-term-rule}, the proposition formation rules in figure~\ref{fig:ert-prop-rule}, and the proof rules and axioms in figures~\ref{fig:ert-proof-rule} and \ref{fig:ert-axiom-rule} respectively.

\subsubsection{Equality}

The heart of the \ert{} type refinement system is the equality proposition \(a =_A b\), which is verified by proofs that \(a = b\) where \(a\) and \(b\) are interpreted as elements of the type \(A\). This has formation rule \hyperref[fig:ert-type-rule]{\brle{Eq-WF}}, which checks that \(a\) and \(b\) are well-typed in an upgraded context. Because equality is a mathematical proposition, we may use ghost variables freely. Furthermore, note that we can consider equalities between terms of any type, including terms of higher type such as function types. This has often been challenging to support with refinement types (see \citet{lh-equality} for a detailed discussion), but is unproblematic with \ert{}.

\begin{figure}
    \begin{gather*}
        \prftree[r]{\rle{Fn-WF}}
            {\Gamma \lrnt A\;\ms{ty}}
            {\Gamma, x: A \lrnt B\;\ms{ty}}
            {\Gamma \lrnt (x: A) \to B\;\ms{ty}} \qquad
        \prftree[r]{\rle{Pair-WF}}
            {\Gamma \lrnt A\;\ms{ty}}
            {\Gamma, x: A \lrnt B\;\ms{ty}}
            {\Gamma \lrnt (x: A) \times B\;\ms{ty}} \\ 
        \prftree[r]{\rle{Intr-WF}}
            {\Gamma \lrnt A\;\ms{ty}}
            {\Gamma, x: A \lrnt B\;\ms{ty}}
            {\Gamma \lrnt \forall x: A, B\;\ms{ty}} \qquad
        \prftree[r]{\rle{Union-WF}}
            {\Gamma \lrnt A\;\ms{ty}}
            {\Gamma, x: A \lrnt B\;\ms{ty}}
            {\Gamma \lrnt \exists x: A, B\;\ms{ty}} \\
        \prftree[r]{\rle{Pre-WF}}
            {\Gamma \lrnt \varphi\;\ms{pr}}
            {\Gamma, u: \varphi \lrnt A\;\ms{ty}}
            {\Gamma \lrnt \lrasm{u}{\varphi}{A}\;\ms{ty}} \qquad
        \prftree[r]{\rle{Set-WF}}
            {\Gamma \lrnt A\;\ms{ty}}
            {\Gamma, x: A \lrnt \varphi\;\ms{pr}}
            {\Gamma \lrnt \{x: A \mid \varphi\}\;\ms{ty}}
        \\
        \prftree[r]{\rle{Unit-WF}}{}
            {\Gamma \lrnt \mb{1}\;\ms{ty}} 
            \qquad
        \prftree[r]{\rle{Coprod-WF}}
            {\Gamma \lrnt A\;\ms{ty}}
            {\Gamma \lrnt B\;\ms{ty}}
            {\Gamma \lrnt A + B\;\ms{ty}} \qquad
        \prftree[r]{\rle{Nats-WF}}{}
            {\Gamma \lrnt \nats\;\ms{ty}}
    \end{gather*}
    \caption{\ert{} Type Well-formedness}
    \Description{
        \ert{} Type Well-formedness
    }
    \label{fig:ert-type-rule}
\end{figure}

\begin{figure}
    \begin{gather*}
        \prftree[r]{\rle{Imp-WF}}
            {\Gamma \lrnt \varphi\;\ms{pr}}
            {\Gamma, u: \varphi \lrnt \psi\;\ms{pr}}
            {\Gamma \lrnt \lrimp{u}{\varphi}{\psi}\;\ms{pr}} \quad
        \prftree[r]{\rle{Or-WF}}
            {\Gamma \lrnt \varphi\;\ms{pr}}
            {\Gamma \lrnt \psi\;\ms{pr}}
            {\Gamma \lrnt \varphi \lor \psi\;\ms{pr}} \quad
        \prftree[r]{\rle{And-WF}}
            {\Gamma \lrnt \varphi\;\ms{pr}}
            {\Gamma, u: \varphi \lrnt \psi\;\ms{pr}}
            {\Gamma \lrnt (u: \varphi) \land \psi\;\ms{pr}} \\
        \prftree[r]{\rle{Univ-WF}}
            {\Gamma \lrnt A\;\ms{ty}}
            {\Gamma, x: A \lrnt \varphi\;\ms{pr}}
            {\Gamma \lrnt \forall x: A, \varphi\;\ms{pr}} \qquad
        \prftree[r]{\rle{Exists-WF}}
            {\Gamma \lrnt A\;\ms{ty}}
            {\Gamma, x: A \lrnt \varphi\;\ms{pr}}
            {\Gamma \lrnt \exists x: A, \varphi\;\ms{pr}} \\
        \prftree[r]{\rle{Eq-WF}}
            {\Gamma \lrnt A\;\ms{ty}}
            {\upg{\Gamma} \lrnt a: A}
            {\upg{\Gamma} \lrnt b: A}
            {\Gamma \lrnt a =_{A} b\;\ms{pr}} \qquad
        \prftree[r]{\rle{True-WF}}{}
            {\Gamma \lrnt \top\;\ms{pr}} \qquad
        \prftree[r]{\rle{False-WF}}{}
            {\Gamma \lrnt \bot\;\ms{pr}}
    \end{gather*}
    \caption{\ert{} Proposition Well-formedness}
    \Description{
        \ert{} Proposition Well-formedness
    }
    \label{fig:ert-prop-rule}
\end{figure}

\begin{figure}
    \begin{gather*}
        \prftree[r]{\rle{Var}}{}
            {\Gamma, x: A \lrnt x: A} \qquad
        \prftree[r]{\rle{Absurd}}
            {\Gamma \lrnt p: \bot}
            {\Gamma \lrnt \ms{absurd}\;p: A} \qquad
        \prftree[r]{\rle{Zero}}
        {\Gamma \lrnt 0: \nats} \qquad
        \prftree[r]{\rle{Succ}}
            {\Gamma \lrnt \ms{succ}: \nats \to \nats}
        \\
        \prftree[r]{\rle{Unit}}
            {\Gamma \lrnt (): \mb{1}} \qquad
        \prftree[r]{\rle{Lam}}
            {\Gamma, x: A \lrnt e: B}
            {\Gamma \lrnt \lambda x: A, e: (x: A) \to B} \qquad
        \prftree[r]{\rle{App}}
            {\Gamma \lrnt l: (x: A) \to B}
            {\Gamma \lrnt r: A}
            {\Gamma \lrnt l\;r: [r/x]B} \\
        \prftree[r]{\rle{Pair}}
            {\Gamma \lrnt l: A}
            {\Gamma \lrnt r: [l/x]B}
            {\Gamma \lrnt (l, r): (x: A) \times B} \qquad
        \prftree[r]{\rle{Inl}}
            {\Gamma \lrnt e: A}
            {\Gamma \lrnt \ms{inl}\;e: A + B} \qquad
        \prftree[r]{\rle{Inr}}
            {\Gamma \lrnt e: B}
            {\Gamma \lrnt \ms{inr}\;e: A + B} \\
        \prftree[r]{\rle{Let-Pair}}
            {\Gamma \lrnt e: (x: A) \times B}
            {\Gamma, z: (x: A) \times B \lrnt C\;\ms{ty}}
            {\Gamma, x: A, y: B \lrnt e': [(x, y)/z]C}
            {\Gamma \lrnt 
                \ms{let}\;(x, y): (x: A) \times B = e 
                \;\ms{in}\; e': [e/z]C} \\
        \prftree[r]{\rle{Cases}}
            {\Gamma, x: A + B \lrnt C\;\ms{ty}}
            {\Gamma \lrnt e: A + B}
            {\Gamma, y: A \lrnt l: [\ms{inl}\;y/x]C}
            {\Gamma, z: B \lrnt r: [\ms{inr}\;z/x]C}
            {\Gamma \lrnt \ms{cases}\;[x \mapsto C]\;e\;(\ms{inl}\;y \mapsto l)\;(\ms{inr}\;z \mapsto r): [e/x]C} \\
        \prftree[r]{\rle{Lam-Pr}}
            {\Gamma, u: \varphi \lrnt e: A}
            {\Gamma \lrnt \lambda u: \varphi.e: \lrasm{u}{\varphi}{A}} \qquad
        \prftree[r]{\rle{App-Pr}}
            {\Gamma \lrnt f: \lrasm{u}{\varphi}{A}}
            {\Gamma \lrnt p: \varphi}
            {\Gamma \lrnt f\;p: [p/x]A} \\
        \prftree[r]{\rle{Set}}
            {\Gamma \lrnt a: A}
            {\Gamma \lrnt p: [a/x]\varphi}
            {\Gamma \lrnt \{a, p\}: \lrset{x}{A}{\varphi}} \qquad
        \prftree[r]{\rle{Pair-Ir}}
            {\upg{\Gamma} \lrnt a: A}
            {\Gamma \lrnt b: [a/x]B}
            {\Gamma \lrnt (\|a\|, b): \exists a: A, B} \\
        \prftree[r]{\rle{Let-Set}}
            {\Gamma \lrnt e: \lrset{x}{A}{\varphi}}
            {\Gamma, z: \lrset{x}{A}{\varphi} \lrnt C\;\ms{ty}}
            {\Gamma, x: A, u: \varphi \lrnt e': [\{x, u\}/z]C}
            {\Gamma \lrnt 
                \ms{let}\;\{x, u\}: \lrset{x}{A}{\varphi} = e 
                \;\ms{in}\; e': [e/z]C} \\
        \prftree[r]{\rle{Lam-Ir}}
            {\Gamma, \|x: A\| \lrnt e: B}
            {\Gamma \lrnt \lambda\|x: A\|, e: \forall x: A, B} \qquad
        \prftree[r]{\rle{App-Ir}}
            {\Gamma \lrnt f: \forall x: A, B}
            {\upg{\Gamma} \lrnt a: A}
            {\Gamma \lrnt f \|a\|: [a/x]B} \\
        \prftree[r]{\rle{Let-Ir}}
            {\Gamma \lrnt e: \exists x: A, B}
            {\Gamma, z: \exists x: A, B \lrnt C\;\ms{ty}}
            {\Gamma, \|x: A\|, y: B \lrnt e': [(\|x\|, y)/z]C}
            {\Gamma \lrnt 
                \ms{let}\;(\|x\|, y): \exists x: A, B = e 
                \;\ms{in}\; e': [e/z]C} \\
        \prftree[r]{\rle{Natrec}}
            {\Gamma, n: \nats \lrnt C\;\ms{ty}}
            {\Gamma \lrnt e: \nats}
            {\Gamma \lrnt z: [0/n]C}
            {\Gamma, \|n: \nats\|, y: C \lrnt s: [\ms{succ}\;n/n]C}
            {\Gamma \lrnt \ms{natrec}[n \mapsto C]\;e\;z\;(\|\ms{succ}\;x\|, y \mapsto s): [e/n]C}
        % \prftree
        %     {\upg{\Gamma} \lrnt a: A}
        %     {\upg{\Gamma} \lrnt b: A} 
        %     {\Gamma \lrnt p: a =_A b}
        %     {\Gamma \lrnt e: [a/x]B}
        %     {\Gamma \lrnt \ms{subst}[x \mapsto B][a][b]\;p\;e: [b/x]B} \qquad
    \end{gather*}
    \caption{\ert{} Term Typing}
    \Description{
        \ert{} Term Typing
    }
    \label{fig:ert-term-rule}
\end{figure}

\begin{figure}
    \begin{gather*}
        \prftree[r]{\rle{Var-Pr}}{}
            {\Gamma, u: \varphi \lrnt u: \varphi} \qquad
        \prftree[r]{\rle{True}}
            {\Gamma \lrnt \langle\rangle: \top} \qquad
        \prftree[r]{\rle{Absurd-Pr}}
            {\Gamma \lrnt p: \bot}
            {\Gamma \lrnt \ms{absurd}\;p: \varphi} \\
        \prftree[r]{\rle{Imp}}
            {\Gamma, u: \varphi \lrnt p: \psi}
            {\Gamma \lrnt \hat{\lambda}u: \varphi, p: \lrimp{u}{\varphi}{\psi}} \qquad
        \prftree[r]{\rle{MP}}
            {\Gamma \lrnt p: \lrimp{u}{\varphi}{\psi}}
            {\Gamma \lrnt q: \psi}
            {\Gamma \lrnt p\;q: [q/u]B} \\
        \prftree[r]{\rle{And}}
            {\Gamma \lrnt p: \varphi}
            {\Gamma \lrnt q: [p/u]\psi}
            {\Gamma \lrnt \langle p, q \rangle: (u: \varphi) \land \psi} \qquad 
        \prftree[r]{\rle{Orl}}
            {\Gamma \lrnt p: \varphi}
            {\Gamma \lrnt \ms{orl}\;p: \varphi \lor \psi} \qquad
        \prftree[r]{\rle{Orr}}
            {\Gamma \lrnt p: \psi}
            {\Gamma \lrnt \ms{orr}\;p: \varphi \lor \psi} \\
        \prftree[r]{\rle{Let-And}}
            {\Gamma \lrnt p: (u: \varphi) \land \psi}
            {\Gamma, w: (u: \varphi) \land \psi \lrnt \theta\;\ms{pr}}
            {\Gamma, u: \varphi, v: \psi \lrnt q: [\langle u, v \rangle/w]\theta}
            {\Gamma \lrnt 
                \ms{let}\;\langle u, v \rangle: (u: \varphi) \land \psi = p 
                \;\ms{in}\; q: [p/w]\theta} \\        
        \prftree[r]{\rle{Cases-Or}}
            {\Gamma, u: \varphi \lor \psi \lrnt \theta\;\ms{pr}}
            {\Gamma \lrnt p: \varphi \lor \psi}
            {\Gamma, v: \varphi \lrnt l: [\ms{orl}\;v/u]\theta}
            {\Gamma, w: \psi \lrnt r: [\ms{orr}\;w/u]\theta}
            {\Gamma \lrnt \ms{cases}_{\ms{or}}\;[u \mapsto \theta]\;p\;(\ms{orl}\;v \mapsto l)\;(\ms{orr}\;w \mapsto r): [e/u]C} \\
        \prftree[r]{\rle{Gen}}
            {\Gamma, \|x: A\| \lrnt p: \varphi}
            {\Gamma \lrnt \hat{\lambda}\|x: A\|, p: \forall x: A, \varphi} \qquad
        \prftree[r]{\rle{Spec}}
            {\Gamma \lrnt p: \forall x: A, \varphi}
            {\upg{\Gamma} \lrnt a: A}
            {\Gamma \lrnt p\;\|a\|: [a/x]\varphi} \\
        \prftree[r]{\rle{Wit}}
            {\upg{\Gamma} \lrnt a: A}
            {\Gamma \lrnt p: [a/x]\varphi}
            {\Gamma \lrnt \langle \|a\|, p \rangle: \exists x: A, \varphi} \\
        \prftree[r]{\rle{Let-Exists}}
            {\Gamma \lrnt p: \exists x: A, \varphi}
            {\Gamma, v: \exists x: A, \varphi \lrnt \psi\;\ms{pr}}
            {\Gamma, x: A, u: \varphi \lrnt q: [\langle \|a\|, p \rangle/v]\psi}
            {\Gamma \lrnt 
                \ms{let}\;\langle \|x\|, u \rangle: \exists x: A, \varphi = p 
                \;\ms{in}\; q: [p/v]\psi} \\
        \prftree[r]{\rle{Let-Pair-Pr}}
            {\upg{\Gamma} \lrnt e: (x: A) \times B}
            {\Gamma, z: (x: A) \times B \lrnt \varphi\;\ms{pr}}
            %{\upg{\Gamma}, x: A, y: B \lrnt e': [(x, y)/z]\varphi}
            {\Gamma, x: A, y: B \lrnt e': [(x, y)/z]\varphi}
            {\Gamma \lrnt 
                \ms{let}\;(x, y): (x: A) \times B = e 
                \;\ms{in}\; e': [e/z]\varphi} \\
        \prftree[r]{\rle{Let-Set-Pr}}
            {\upg{\Gamma} \lrnt e: \lrset{x}{A}{\varphi}}
            {\Gamma, z: \lrset{x}{A}{\varphi} \lrnt \psi\;\ms{pr}}
            %{\upg{\Gamma}, x: A, u: \varphi \lrnt e': [\{x, y\}/z]\psi}
            {\Gamma, x: A, u: \varphi \lrnt e': [\{x, y\}/z]\psi}
            {\Gamma \lrnt 
                \ms{let}\;\{x, u\}: \lrset{x}{A}{\varphi} = e 
                \;\ms{in}\; e': [e/z]\psi} \\
        \prftree[r]{\rle{Let-Ir-Pr}}
            {\upg{\Gamma} \lrnt e: \exists x: A, \varphi}
            {\Gamma, z: \exists x: A, B \lrnt \varphi\;\ms{ty}}
            %{\upg{\Gamma}, x: A, y: B \lrnt e': [(\|x|\|, y)/z]\varphi}
            {\Gamma, x: A, y: B \lrnt e': [(\|x|\|, y)/z]\varphi}
            {\Gamma \lrnt 
                \ms{let}\;(\|x\|, y): \exists x: A, B = e 
                \;\ms{in}\; e': [e/z]\varphi} \\
        \prftree[r]{\rle{Subst}}
            {\upg{\Gamma} \lrnt a: A}
            {\upg{\Gamma} \lrnt b: A} 
            {\Gamma \lrnt p: a =_A b}
            {\Gamma \lrnt q: [a/x]\varphi}
            {\Gamma \lrnt \ms{subst}[x \mapsto \varphi][a][b]\;p\;q: [b/x]\varphi} \\
        \prftree[r]{\rle{Cases-Pr}}
            {\Gamma, x: A + B \lrnt \varphi;\ms{pr}}
            {\Gamma \lrnt e: A + B}
            %{\upg{\Gamma}, y: A \lrnt l: [\ms{inl}\;y/x]\varphi}
            %{\upg{\Gamma}, z: B \lrnt r: [\ms{inr}\;z/x]\varphi}
            {\Gamma, y: A \lrnt l: [\ms{inl}\;y/x]\varphi}
            {\Gamma, z: B \lrnt r: [\ms{inr}\;z/x]\varphi}
            {\Gamma \lrnt \ms{cases}\;[x \mapsto \varphi]\;e\;(\ms{inl}\;y \mapsto l)\;(\ms{inr}\;z \mapsto r): [e/x]\varphi} \\
        \prftree[r]{\rle{Ind}}
            {\Gamma, n: \nats \lrnt \varphi\;\ms{pr}}
            {\upg{\Gamma} \lrnt e: \nats}
            % {\upg{\Gamma} \lrnt z: [0/n]\varphi}
            % {\upg{\Gamma}, x: \nats, y: [x/n]\varphi: s: [\ms{succ}\;x/n]\varphi}
            {\Gamma \lrnt z: [0/n]\varphi}
            {\Gamma, n: \nats, y: \varphi: s: [\ms{succ}\;n/n]\varphi}
            {\Gamma \vdash \ms{ind}[n \mapsto \varphi]\;e\;z\;(\ms{succ}\;n, y \mapsto s): [e/n]\varphi}
    \end{gather*}
    \caption{\ert{} Proof Typing}
    \Description{
        \ert{} Proof Typing
    }
    \label{fig:ert-proof-rule}
\end{figure}

\begin{figure}
    \begin{gather*}
        \prftree[r]{\rle{Rfl}}
            {\upg{\Gamma} \lrnt a: A}
            {\Gamma \lrnt \ms{rfl}\;a: a =_A a} \qquad 
        \prftree[r]{\rle{Uniq}}
            {\upg{\Gamma} \lrnt a: \mb{1}}
            {\Gamma \lrnt \ms{uniq}\;a: a =_{\mb{1}} ()} \\
        \prftree[r]{\rle{Discr}}
            {\upg{\Gamma} \lrnt a: A}
            {\upg{\Gamma} \lrnt b: B}
            {\Gamma \lrnt p: \ms{inl}\;a =_{A + B} \ms{inr}\;b}
            {\Gamma \lrnt \ms{discr}\;a\;b\;p: \bot}
            \\
        \prftree[r]{\rle{\(\beta_{\ms{pr}}\)}}
            {\upg{\Gamma}, u: \varphi \lrnt e: A}
            {\Gamma \lrnt p: \varphi}
            {\Gamma \lrnt 
                \beta_{\ms{pr}}(u \mapsto e)\;p:
                (\hat{\lambda}u: \varphi.e)\;p =_{[p/u]A} [p/u]e
            } \\
        \prftree[r]{\rle{\(\beta_{\ms{ty}}\)}}
            {\upg{\Gamma}, x: A \lrnt e: B}
            {\upg{\Gamma} \lrnt a: A}
            {\Gamma \lrnt 
                \beta_{\ms{ty}}(x \mapsto e)\;a:
                (\lambda x: A.e)\;a =_{[a/x]B} [a/x]e
            } \\
        \prftree[r]{\rle{\(\beta_{\ms{ir}}\)}}
            {\upg{\Gamma}, x: A \lrnt e: B}
            {\upg{\Gamma} \lrnt a: A}
            {\Gamma \lrnt 
                \beta_{\ms{ir}}(x \mapsto e)\;a:
                (\lambda\|x: A\|.e)\;\|a\| =_{[a/x]B} [a/x]e
            } \\
        \prftree[r]{\rle{\(\beta_{\ms{left}}\)}}
            {\upg{\Gamma} \lrnt a: A}
            {\upg{\Gamma}, y: A \lrnt l: [\ms{inl}\;y/x]C}
            {\upg{\Gamma}, z: B \lrnt r: [\ms{inr}\;z/x]C}
            {
                \begin{array}{rl}
                \Gamma \lrnt&
                    \beta_{\ms{left}}[x \mapsto C]
                    \;(\ms{inl}\;y \mapsto l)
                    \;(\ms{inr}\;z \mapsto r)
                    \;(\ms{inl}\;a)
                \\:& \ms{cases}[x \mapsto C]\;(\ms{inl}\;a)\;(\ms{inl}\;y \mapsto l)\;(\ms{inr}\;z \mapsto r) =_{[\ms{inl}\;a/x]C} [a/y]l
                \end{array}
            } \\
        \prftree[r]{\rle{\(\beta_{\ms{right}}\)}}
            {\upg{\Gamma} \lrnt b: B}
            {\upg{\Gamma}, y: A \lrnt l: [\ms{inl}\;y/x]C}
            {\upg{\Gamma}, z: B \lrnt r: [\ms{inr}\;z/x]C}
            {
                \begin{array}{rl}
                    \Gamma \lrnt& 
                        \beta_{\ms{right}}[x \mapsto C]
                        \;(\ms{inl}\;y \mapsto l)
                        \;(\ms{inr}\;z \mapsto r)
                        \;(\ms{inr}\;b)
                    \\:& \ms{cases}[x \mapsto C]
                        \;(\ms{inr}\;b)
                        \;(\ms{inl}\;y \mapsto l)
                        \;(\ms{inr}\;z \mapsto r) 
                    =_{[\ms{inr}\;b/x]C} [b/z]r
                \end{array}
            } \\
        \prftree[r]{\rle{\(\beta_{\ms{zero}}\)}}
            {\upg{\Gamma} \lrnt z: [0/x]C}
            {\upg{\Gamma}, x: \nats, y: C \lrnt s: [\ms{succ}\;x/x]C}
            {
                \Gamma \lrnt 
                    \beta_{\ms{zero}}[x \mapsto C]
                    \;z
                    \;(\|\ms{succ}\;x\|, y \mapsto s)
                : \ms{natrec}[x \mapsto C]
                    \;0
                    \;z
                    \;(\|\ms{succ}\;x\|, y \mapsto s)
                =_{[0/x]C} z
            } \\
        \prftree[r]{\rle{\(\beta_{\ms{succ}}\)}}
            {\upg{\Gamma} \lrnt e: \nats}
            {\upg{\Gamma} \lrnt z: [0/x]C}
            {\upg{\Gamma}, x: \nats, y: C \lrnt s: [\ms{succ}\;x/x]C}
            {
                \begin{array}{rl}
                    \Gamma\lrnt & 
                        \beta_{\ms{succ}}[x \mapsto C]
                        \;(\ms{succ}\;e)
                        \;z
                        \;(\|\ms{succ}\;x\|, y \mapsto s)
                    \\:& \ms{natrec}[x \mapsto C]
                        \;(\ms{succ}\;e)
                        \;z
                        \;(\|\ms{succ}\;x\|, y \mapsto s)
                    \\&=_{[\ms{succ}\;e/x]C} [(\ms{natrec}[x \mapsto C]
                        \;e
                        \;z
                        \;(\|\ms{succ}\;x\|, y \mapsto s))/y
                        ][e/x]s
                \end{array}
            } \\
        \prftree[r]{\rle{\(\beta_{\ms{pair}}\)}}
            {\upg{\Gamma} \lrnt a: A}
            {\upg{\Gamma} \lrnt b: [a/y]B}
            {\upg{\Gamma}, y: A, z: B \lrnt e: [(y, z)/x]C}
            {
                \Gamma \lrnt 
                    \beta_{\ms{pair}}
                    \;(a, b)
                    \;((y, z) \mapsto e)
                : (\ms{let}\;(y, z) = (a, b)\;\ms{in}\;e) =_{[(a, b)/x]C}
                    [b/z][a/y]e
            } \\
        \prftree[r]{\rle{\(\beta_{\ms{set}}\)}}
            {\upg{\Gamma} \lrnt a: A}
            {\upg{\Gamma} \lrnt p: [a/y]\varphi}
            {\upg{\Gamma}, y: A, u: \varphi \lrnt e: [\{y, u\}/x]C}
            {
                \Gamma \lrnt 
                    \beta_{\ms{set}}
                    \;\{a, p\}
                    \;((y, u) \mapsto e)
                : (\ms{let}\;\{y, u\} = \{a, p\}\;\ms{in}\;e) =_{[\{a, p\}/x]C}
                    [p/u][a/y]e
            } \\
        \prftree[r]{\rle{\(\beta_{\ms{repr}}\)}}
            {\upg{\Gamma} \lrnt a: A}
            {\upg{\Gamma} \lrnt b: [a/y]B}
            {\upg{\Gamma}, y: A, z: B \lrnt e: [(\|y\|, z)/x]C}
            {
                \Gamma \lrnt 
                    \beta_{\ms{repr}}
                    \;(\|a\|, b)
                    \;((y, z) \mapsto e)
                : (\ms{let}\;(\|y\|, z) = (\|a\|, b)\;\ms{in}\;e) =_{[(\|a\|, b)/x]C}
                    [b/z][a/y]e
            } \\
        \prftree[r]{\rle{\(\eta_{\ms{ty}}\)}}
            {\upg{\Gamma} \lrnt f: (x: A) \to B}
            {\Gamma \lrnt \eta_{\ms{ty}}\;f: \lambda x: A, f\;x =_{(x: A) \to B} f} \qquad
        \prftree[r]{\rle{Ir-Pr}}
            {\Gamma \vdash A\;\ms{ty}}
            {\upg{\Gamma}, u: \varphi \lrnt e: A}
            {\upg{\Gamma} \lrnt p: \varphi}
            {\upg{\Gamma} \lrnt q: \varphi}
            {\Gamma \lrnt \ms{ir}_{\ms{pr}}[u \mapsto e]\;p\;q: [p/u]e =_A [q/u]e}
            \\
        \prftree[r]{\rle{Ir-Ty}}
            {\Gamma \vdash B\;\ms{ty}}
            {\upg{\Gamma} \lrnt e: \forall x: A, B}
            {\upg{\Gamma} \lrnt a: A}
            {\upg{\Gamma} \lrnt b: A}
            {\Gamma \lrnt \ms{ir}_{\ms{ty}}\;e\;a\;b: e\;\|a\| =_B e\;\|b\|} \\
        \prftree[r]{\rle{$\eta_{\ms{ir}}$}}
            {\upg{\Gamma} \lrnt f: \forall x: A, B}
            {\upg{\Gamma} \lrnt g: \forall x: A, B}
            {\upg{\Gamma} \lrnt i: A}
            {\Gamma, x: A \lrnt p: f\;\|x\| =_B g\;\|x\|}
            {\Gamma \lrnt \eta_{\ms{ir}}\;f\;g\;i\;p: f =_{\forall x: A, B} g} \\
        \prftree[r]{\rle{$\eta_{\ms{pr}}$}}
            {\upg{\Gamma} \lrnt f: \lrasm{x}{\varphi}{B}}
            {\upg{\Gamma} \lrnt g: \lrasm{x}{\varphi}{B}}
            {\upg{\Gamma} \lrnt i: \varphi}
            {\Gamma, x: \varphi \lrnt p: f\;x =_B g\;x}
            {\Gamma \lrnt \eta_{\ms{pr}}\;f\;g\;i\;p: f =_{\lrasm{x}{\varphi}{B}} g}
    \end{gather*}
    \caption{\ert{} Axiom Typing}
    \Description{
        \ert{} Axiom Typing
    }
    \label{fig:ert-axiom-rule}
\end{figure}

The introduction rule is the standard reflexivity axiom, \hyperref[fig:ert-axiom-rule]{\brle{Rfl}}, and  equalities may be eliminated via substitution using the rule \hyperref[fig:ert-proof-rule]{\brle{Subst}}. The substitution eliminator is powerful enough to prove the other equality axioms, such as, for example, transitivity: 
% \begin{equation}
%     \prftree
%         {\Gamma \lrnt p: a=_Ab}
%         {\ms{symm}\;p \equiv
%             \ms{subst}[x \mapsto x =_Aa][a][b]\;p\;(\ms{rfl}\;a)
%          : b =_A a}
% \end{equation}
\begin{equation}
    \prftree
        {\Gamma \lrnt p: a=_Ab}
        {\Gamma \lrnt q: b=_Ac}
        {\ms{trans}\;p\;q \equiv
            \ms{subst}[x \mapsto a =_Ax][b][c]\;q\;p
         : a =_A c}
\end{equation}
% \begin{equation}
%     \prftree
%         {\Gamma \lrnt p: f=_{(x: A) \to B}g}
%         {\upg{\Gamma} \lrnt a: A}
%         {\ms{congr}_{\ms{left}}\;p\;a \equiv
%             \ms{subst}[y \mapsto y\;a=_{[a/x]B}y\;b][f][g]\;p\;
%                 (\ms{rfl}\;f)
%          : f\;a =_{[a/x]B} g\;a}
% \end{equation}
Type equality \ert{} is just \(\alpha\)-equivalence, and so any equations which would have come via judgemental equality in a dependent type theory must be expressed as  equality axioms. For example, beta-reduction for functions is expressed via the axiom \hyperref[fig:ert-axiom-rule]{\(\beta_{\ms{ty}}\)}, and there are similar rules for each of the type constructors in the language. Furthermore, because proofs are computationally irrelevant, they support an extensionality principle: the axiom \hyperref[fig:ert-axiom-rule]{\brle{Ir-Pr}} lets us replace any proof \(p\) with any other proof \(q\).

\subsubsection{Type Structure}
\ert{} has (refinements of) the usual type constructors of the simply-typed lambda calculus, such as functions, pairs, sum types, and datatypes like natural numbers, as well as type constructors specific to refinement types such as subset types, generalised intersections and unions, and preconditions.

Dependent functions of type \((x:A) \to B\) may be introduced by lambda abstraction, via the rule \hyperref[fig:ert-term-rule]{\brle{Lam}}, and may be eliminated by application, via the rule \hyperref[fig:ert-term-rule]{\brle{App}}. The corresponding \(\beta\) and \(\eta\) equations are introduced axiomatically, with the rules \hyperref[fig:ert-axiom-rule]{\(\beta_{\ms{ty}}\)} and \hyperref[fig:ert-axiom-rule]{\(\eta_{\ms{ty}}\)}.
Each of the \(\beta\) and \(\eta\) axioms is subscripted with an annotation naming the type former it is for. For example, the subscript "\(\ms{ty}\)" in \(\beta_{\ms{ty}}\) refers to the fact that dependent function types are parameterized by a term variable (with a type). The rule application itself is annotated with the function body and argument. 

While dependent function types abstract over a \emph{computational} variable, we can also abstract over \emph{ghost} or \emph{(computationally) irrelevant} variables, yielding a form of \emph{intersection type}. The type well-formedness conditions for \(\forall x:A, B\) (in \hyperref[fig:ert-type-rule]{\brle{Intr-WF}}) are essentially the same conditions as for dependent functions, but the introduction rule \hyperref[fig:ert-term-rule]{\brle{Lam-Ir}} checks the body assuming the parameter is irrelevant.  

We can eliminate a term of intersection type by applying it to an expression which is well-typed in the upgraded context \(\upg{\Gamma}\), i.e., which may contain ghost variables, via the rule \hyperref[fig:ert-term-rule]{\brle{App-Ir}}. Similarly to the case for dependent functions, reduction must be encoded as an axiom \hyperref[fig:ert-axiom-rule]{\(\beta_{\ms{ir}}\)}, where the "\(\ms{ir}\)" stands for "irrelevant."
We may also introduce an \emph{irrelevance axiom}, \hyperref[fig:ert-axiom-rule]{\brle{Ir-Ty}}, which essentially says that ghost arguments do not matter for the purposes of determining equality whenever the ghost variable does not occur in the result type. 

% Note that, for reasons of typability, the result type \(B\) of the intersection type \(\forall x: A, B\) is here required to be independent of \(x\), i.e., \(B\) must be well typed in \(\Gamma\) (rather than \(\Gamma, x: A\)).

We move on to introduce dependent pair types with the type formation
rule \hyperref[fig:ert-type-rule]{\brle{Pair-WF}}.  The introduction
rule \hyperref[fig:ert-term-rule]{\brle{Pair}} looks a bit like the
introduction rule for sigma-types in dependent type theories, with the
type of the second component varying according to the first component,
and both components computationally relevant.

The elimination form is a let-binding form (in \hyperref[fig:ert-term-rule]{\brle{Let-Pair}}). We may also eliminate into proofs using \hyperref[fig:ert-proof-rule]{\brle{Let-Pair-Pr}}; note that, in this case, the expression \(e: (a: A) \times B\) may contain ghost variables (as it only needs to be well-typed in \(\upg{\Gamma}\)). As before, reduction for dependent pair elimination must be encoded as an axiom, \hyperref[fig:ert-axiom-rule]{\(\beta_{\ms{pair}}\)}.

Often, however, we may want to be able to consider, dually to intersection types, \emph{union types} \(\exists x: A, B(x)\) (\hyperref[fig:ert-type-rule]{\brle{Union-WF}}), which we may views as dependent pairs conditioned on a ghost variable, or, set-theoretically, as elements of \(B(a)\) for some valid \(a\). Similarly to for dependent pairs, we support an introduction rule \hyperref[fig:ert-term-rule]{\brle{Pair-Ir}}, and elimination via let-binding using rules \hyperref[fig:ert-term-rule]{\brle{Let-Ir}} and \hyperref[fig:ert-proof-rule]{\brle{Let-Ir-Pr}}. Note that, unlike for dependent pairs, and similarly to intersection types, \(a\) only needs to be well-typed in \(\upg{\Gamma}\) (rather than \(\Gamma\)) in the introduction rule \hyperref[fig:ert-term-rule]{\brle{Pair-Ir}}, while in \hyperref[fig:ert-term-rule]{\brle{Let-Ir}}, the binder \(\|x: A\|\) is a ghost binder rather than a term binder. Finally, as before, we also introduce a reduction rule for let-bindings, \hyperref[fig:ert-axiom-rule]{\(\beta_{\ms{ir}}\)}.

Just as dependent functions and pairs have corresponding type formers quantifying over ghost variables rather than term variables, we may also construct type formers predicated over propositions. In particular, we may consider the \emph{precondition type}: essentially a closure yielding an element of the type \(A\) if the proposition \(p\) is true; this has introduction rule \hyperref[fig:ert-type-rule]{\brle{Pre-WF}}. Introduction is by abstracting a term over a proof variable, with rule \hyperref[fig:ert-term-rule]{\brle{Lam-Pr}}.
Note that the type \(A\) in \hyperref[fig:ert-term-rule]{\brle{Lam-Pr}} is allowed to depend on a proof \(u: \varphi\); this is because we may consider precondition types \(\lrasm{u}{\varphi}{A}\) in which \(A\) is only well-formed if \(\varphi\) holds. For example, consider a function \(f: \{X \mid \varphi(x)\} \to X\) (we will cover subset types shortly); to reason about values of \(f\) for \(a: X\), we need \(\varphi(a)\) to hold, hence, we require dependency on proofs to be able to type
\(\lrasm{u}{\varphi(a)}{\{z: \{y: X \mid \varphi(y)\} \mid f\;z = f\;\{a, u\}\}}\).
Similarly to for the computational and ghost variable cases, we must also introduce a reduction rule \hyperref[fig:ert-axiom-rule]{\(\beta_{\ms{pr}}\)}.
Dually, we may introduce the \emph{subset type} former
\hyperref[fig:ert-type-rule]{\brle{Set-WF}},
representing, in essence, elements \(a\) of type \(A\) satisfying the predicate \(\varphi(a)\). This has the expected introduction rule
\hyperref[fig:ert-term-rule]{\brle{Set}}
and supports elimination via let-binding with rules
\hyperref[fig:ert-term-rule]{\brle{Let-Set}} and 
\hyperref[fig:ert-proof-rule]{\brle{Let-Set-Pr}}. We also introduce a reduction rule, \hyperref[fig:ert-axiom-rule]{\(\beta_{\ms{set}}\)}, as expected.

Currently, we have the ability to manipulate data and associate it
with propositions, but do not yet have any bona fide data types. While
there is no semantic obstacle to introducing a full language of
datatype declarations, for simplicity, we will restrict ourselves to
the unit type, sum types, and the natural numbers. Other inductive types
follow a similar pattern. 
As usual, a value of the unit type may be introduced with rule
\hyperref[fig:ert-term-rule]{\brle{Unit}}; rather than the eliminator
we would expect from dependent type theory, we may simply couple this
with an axiom, \hyperref[fig:ert-axiom-rule]{\brle{Uniq}}, stating
that every member of the unit type is equal to \(()\). While on it's
own this is not a particularly interesting datatype, when combined
with coproduct types \(A + B\), we may define, for example, the type
of Booleans as \(\mb{2} \equiv \mb{1} + \mb{1}\), allowing us to
construct finite types. Coproducts may be introduced via injection
(via rules \hyperref[fig:ert-term-rule]{\brle{Inl}} and
\hyperref[fig:ert-term-rule]{\brle{Inr}}) and eliminated by case
splitting (via rules \hyperref[fig:ert-term-rule]{\brle{Cases}} and
\hyperref[fig:ert-proof-rule]{\brle{Cases-Pr}}). To demonstrate
support for infinite types, we introduce the natural numbers
\(\nats\), constants of which may be built up using
\hyperref[fig:ert-term-rule]{\brle{Zero}} and
\hyperref[fig:ert-term-rule]{\brle{Succ}}. The elimination rule,
\hyperref[fig:ert-term-rule]{\brle{Natrec}}, implements essentially
iteration with the current step available as a \emph{ghost} variable
(for reasoning about in propositions); it's computational semantics
are given by the axioms
\hyperref[fig:ert-axiom-rule]{\(\beta_{\ms{zero}}\)} and
\hyperref[fig:ert-axiom-rule]{\(\beta_{\ms{succ}}\)}. Furthermore,
just as we may construct terms recursively via \(\ms{natrec}\), we may
also perform proofs by induction via \(\ms{ind}\) using rule
\hyperref[fig:ert-proof-rule]{\brle{Ind}}. Note that, unlike in
\hyperref[fig:ert-term-rule]{\brle{Natrec}}, the expression \(n\) over
which we are performing induction is allowed to contain ghost
variables.

\subsubsection{Propositional Structure}

Now that we have given essentially a complete description of the \ert{}'s terms and their computational behaviour, we can describe the main components of \ert{}'s propositional logic, which for the most part mirror the term formers, since we encode proofs in first-order logic as \(\lambda\)-terms via the Curry-Howard correspondence. We begin by introducing propositions \(\top\) and \(\bot\); the former is only equipped with introduction rule \hyperref[fig:ert-proof-rule]{\rle{True}}, whereas \(\bot\), being an initial object, is only equipped with the elimination rules \hyperref[fig:ert-proof-rule]{Absurd-Pr} and \hyperref[fig:ert-term-rule]{Absurd}. The latter rule is especially important, as it is the main way with which our logic is capable of interacting with our term calculus by allowing us to safely erase unreachable branches from, e.g., a \(\ms{natrec}\) or \(\ms{cases}\) expression.

Proofs of equality do not interact meaningfully with the term calculus, since they are logical formulae, and our logic is classical and nonconstructive. So principles such as unique choice (which permit turning a proof that there exists a unique element of type $A$ into a computational value of type $A$) are not valid. The only case in which this is allowable is for the empty type (see the discussion of the \hyperref[fig:ert-term-rule]{Absurd} rule above). To make use of this, we need the additional axiom \hyperref[fig:ert-axiom-rule]{\brle{Discr}}, which essentially says that the right-hand and left-hand side of a coproduct type are disjoint. This suffices to effectively introduce disequality into our type theory, as desired.

We may now introduce the rest of the connectives of first-order logic, suitably modified in order to fit our setting. In particular, we have \emph{implication} \(\lrimp{u}{\varphi}{\psi}\), which, as per the Curry-Howard correspondence, is introduced by abstracting over a proposition variable, via rule \hyperref[fig:ert-proof-rule]{\brle{Imp}}, and eliminated via modus ponens (under Curry-Howard, application) \hyperref[fig:ert-proof-rule]{\brle{MP}}. Similarly, we have \emph{conjunction} \((u: \varphi) \land \psi\), which is introduced by constructing a pair, via rule \hyperref[fig:ert-proof-rule]{\brle{And}}, and eliminated via a let-binding, with rule \hyperref[fig:ert-proof-rule]{\brle{Let-And}}. We note that both associated proposition formers, \hyperref[fig:ert-prop-rule]{\brle{Imp-WF}} and \hyperref[fig:ert-prop-rule]{\brle{And-WF}}, are "dependent," in that their right-hand side \(\psi\) is allowed to depend on a proof variable \(u\) for the left-hand side \(\varphi\). This is because we allow the case where \(\psi\) is not well-formed without \(\varphi\) holding (for example, because it is about a term that requires a proof of \(\varphi\)). On the other hand, similarly to for coproducts, the rules for disjunction \(\varphi \lor \psi\) are simpler, with introduction by injection via rules \hyperref[fig:ert-proof-rule]{\brle{Orl}} and \hyperref[fig:ert-proof-rule]{\brle{Orr}}, and elimination via case splitting with rule \hyperref[fig:ert-proof-rule]{\brle{Cases-Or}}.

Finally, just as we may consider a type quantifying over a proposition, to support full first-order logic, we must also be able to consider propositions quantified over types. In particular, we may introduce \emph{universally quantified} propositions with formation rule \hyperref[fig:ert-prop-rule]{\brle{Univ-Wf}}.
These may be introduced by generalization via rule \hyperref[fig:ert-proof-rule]{\brle{Gen}}, and specialized via elimination rule \hyperref[fig:ert-proof-rule]{\brle{Spec}}. Similarly, we may introduce \emph{existentially quantified} propositions with formation rule \hyperref[fig:ert-prop-rule]{\brle{Exists-WF}}.
We introduce proofs of an existentially quantified proposition by introducing a witness via rule \hyperref[fig:ert-proof-rule]{\brle{Wit}}, and may eliminate proofs via let-binding with rule \hyperref[fig:ert-proof-rule]{\brle{Let-Exists}}. Note in particular that, in both cases, we treat the variable being quantified over as a \emph{ghost} variable, since it is appearing in a proposition. Furthermore, since propositions have no computational semantics, a reduction rule is unnecessary for either.

Altogether, our $\beta$ and $\eta$ rules are quite powerful, enabling us to prove more
extensionality properties than a casual look at the axioms may suggest. In particular,
fixing a term \(e\) with free variable \(z: \{x: A \mid \varphi\}\), the $\eta$-rule for subsets can be written as:
\begin{equation}
    \begin{array}{rl}
    \eta_{\ms{set}}[C] & \equiv \hat{\lambda}\|a: \{x: A \mid \varphi\}\|, \ms{let}\;\{y, v\} = a\;\ms{in}\;
        \beta_{\ms{set}}\;\{y, v\}\;((x, u) \mapsto ([\{x, u\}/z]e))
    \\ &: \forall a: \{x: A \mid \varphi\}, \ms{let}\;\{x, u\} = a\;\ms{in}\;[\{x, u\}/z]e = [a/z]e
    \end{array}
\end{equation}
(note that \(a, x, u, y, v\) are assumed to be fresh variables here). In general, we only need to introduce explicit extensionality axioms for intersection types (\hyperref[fig:ert-proof-rule]{\brle{$\eta_{\ms{ir}}$}}) and precondition types (\hyperref[fig:ert-proof-rule]{\brle{$\eta_{\ms{pr}}$}}), with all the other $\eta$ and extensionality rules we would expect to hold being derivable from the rest of the axioms, with the sole exception of function extensionality, which is not compatible with our current semantics. See section~\ref{ssec:semantics} for the explanation of why. 

Overall, \ert{}'s logic is essentially multi-sorted first-order logic, with the sorts drawn from the types of \ert{}'s programming language. Since \ert{} has function types, this means that the \ert{} logic is fairly close to $\mathrm{PA}^\omega$, Peano arithmetic over the full type hierarchy. So any property provable about \ert{} terms with first-order logic and induction should be provable. (We have not proved any theorems about the expressivity of \ert{}'s logic, though.)

\subsection{Syntactic Metatheory}
\label{ssec:metatheory}

\ert{} satisfies the expected syntactic properties of substitution and regularity.
To show this, we define substitutions as functions \(\sigma: \ms{Var} \to \ms{Term} \uplus \ms{Proof}\), and can recursively define capture-avoiding substitution on terms/proofs \(e\), written \([\sigma]e\), in the obvious way.
We define a well-formed substitution from \(\Gamma\) to \(\Delta\), written \(\Gamma \lrnt' \sigma: \Delta\), as a substitution satisfying the following conditions: 
\begin{equation}
    \begin{aligned}
        [(x: A) \in \Gamma] &\implies \Delta \lrnt \sigma\;x: [\sigma]A \\
        [(u: \varphi) \in \Gamma] &\implies \Delta \lrnt \sigma\;u: [\sigma]\varphi \\
        [\|x: A\| \in \Gamma] &\implies [\|\sigma\;x: [\sigma]A\| \in \Delta] \lor [\Delta \lrnt \sigma\;x: [\sigma]A]
    \end{aligned}
\end{equation}
Furthermore, we say \(\sigma\) is a \emph{strict} substitution, written \(\Gamma \lrnt \sigma: \Delta\), if ghost variables in \(\Gamma\) are only replaced with ghost variables in \(\Delta\).
\begin{lemma}[Syntactic Substitution]
    If \(\Gamma \lrnt' \sigma: \Delta\) and \(\Gamma \lrnt a: A\), 
    then \(\Delta \lrnt [\sigma]a: [\sigma]A\).
    \label{lem:subst}

    \formalization{LogicalRefinement/Typed/Subst.lean}{theorem HasType.subst'} 
\end{lemma}
\noindent
The proof of substitution is a routine induction, which as usual requires first proving weakening. 
Once we know that substitution holds, we can prove regularity: 
\begin{lemma}[Syntactic Regularity]
    If \(\Gamma \lrnt a: A\), then \(\Gamma \lrnt A\;\ms{ty}\). 
    Also, if \(\Gamma \lrnt p: \varphi\), then \(\Gamma \lrnt \varphi\;\ms{pr}\).
    \label{lem:regularity}

    \formalization{LogicalRefinement/Typed/Regular.lean}{theorem HasType.regular}
\end{lemma}
\noindent This requires syntactic substitution since some of the typing rules (such as \hyperref[fig:ert-term-rule]{\brle{App}}) involve a substitution in the result types. 
One other result we will use later is that substitutions can be upgraded; that is, \(\Gamma \lrnt \sigma: \Delta \implies \upg{\Gamma} \lrnt \sigma: \upg{\Delta}\). To avoid confusion, we  write the latter as \(\upg{\Gamma} \lrnt \upg{\sigma}: \upg{\Delta}\), with the upgrade on substitutions taken to be the identity.

\section{Semantics}

\label{sec:semantics}

To give a denotational semantics for the \ert{} calculus, we first
show that all the proofs and dependencies in an \ert{} term can be
erased in a compositional way. This yields a simple type for each
\ert{} type, and a simply-typed term for each \ert{} term. We then
give a semantics for each \ert{} type as a subset of the denotational
semantics of the erasure for that type. Finally, we show that each
well-typed \ert{} term lies in the subset defined by its type.

In section~\ref{ssec:stlc}, we recall the syntax and semantics of the
simply-typed lambda calculus. In section~\ref{ssec:erasure}, we give
an erasure function from \ert{} types and terms to \stlc{} types and
terms respectively, and prove some expected properties like
preservation of well-typedness and semantic substitution (where the
denotational semantics of an \ert{} term are taken to be the semantics
of its erasure). Finally, in section~\ref{ssec:semantics}, we give a
semantics to \ert{} types by assigning each a subset, and show that
the denotations of all well-typed \ert{} terms lie in the subset
assigned to their type, i.e., semantic regularity. From this, we
deduce that "well typed programs don't go wrong".

For the more categorically-minded reader, we interpret \stlc{} in the
Kleisli category of the exception monad $M(X)$ (in the category of sets), and then interpret \ert{} in
terms of the subset fibration over it. That is, an \ert{} type can be
understood as a pair $(X, P \subseteq X)$, where $X$ is a type of
\stlc{}, and $P$ is a predicate on that type; and \ert{} terms are
maps $f \in (\Gamma, P_\Gamma) \to (M(X), P_{M(X)})$ such that for all
$\gamma \in P_\Gamma$, we have $f(\gamma) \in P_{M(X)}$. Semantically, our
erasure operation amounts to the forgetful functor into the Kleisli
category which drops the
property information.

\subsection{The Simply-Typed Lambda Calculus}
\label{ssec:stlc}

We begin by providing typing rules for \stlc{} in figure~\ref{fig:stlc-typing} (a formal grammar is provided in figure~\ref{fig:stlc-grammar} in the appendix).  This is a standard lambda calculus with functions, sums, products, natural numbers, as well as a simple effect: error stops. We define \(\tmstlc\) to be the set of \stlc{} terms. Similarly to the \ert{} calculus, given a function \(\sigma: \ms{Var} \to \tmstlc\), we may then recursively define (capture-avoiding) substitution of a term \(t\) in the usual manner. We say that \(\sigma\) is a substitution from \(\Gamma\) to \(\Delta\), written \(\Gamma \stnt \sigma: \Delta\), if it satisfies the property that \((a: A) \in \Gamma \implies \Delta \stnt \sigma\;a: A\).  

We may then state the usual property of syntactic substitution as follows: 
\begin{lemma}[Syntactic Substitution (\stlc)]
    Given a substitution \(\Gamma \stnt \sigma: \Delta\) and \(\Gamma \stnt t: A\), we have \(\Delta \stnt [\sigma]t: A\)

    \formalization{LogicalRefinement/Stlc/Basic.lean}{theorem Stlc.HasType.subst}
\end{lemma}
% Note that, unlike for \ert{}, there is no need to prove syntactic regularity since every type produced by the grammar in figure~\ref{fig:stlc-grammar} is well-formed.
We may now give \stlc{} a denotational semantics. Fixing the exception monad \(\ms{M}\), with exception \(\ms{error}\), we begin by giving denotations for \stlc{} types in figure~\ref{fig:stlc-ty-denot}, using Moggi's call-by-value semantics for types~\cite{moggi91}. We may then define the denotation of an \stlc{} context elementwise by taking \(\dnt{\cdot} = \mb{1}\), \(\dnt{\Gamma, x: A} = \dnt{\Gamma} \times \ms{M}\;\dnt{A}\).
Despite the fact that our semantics is call-by-value, the interpretation of each hypothesis lives in the monad \(\ms{M}\). We do this so our denotational semantics can interpret substituting arbitrary terms for variables, and not just values for variables. (Since the substitution rule permits substituting arbitrary terms for variables, our semantics has to support this too.)
For each variable \(x: A\) in a context \(\Gamma\), we define pointwise projections \(\pi_x: \dnt{\Gamma} \to \dnt{A}\). We define (in figure~\ref{fig:stlc-term-denot}) the denotation of a derivation \(\Gamma \stnt a: A\) as a function of type \(\dnt{\Gamma} \to \ms{M}(\dnt{A})\), which takes environments (elements of \(\dnt{\Gamma}\)) to elements of the monadic type \(\ms{M}(\dnt{A})\). 

%% Denotation of substitutions

We can now give a relatively straightforward account of the denotational semantics of an \stlc{} substitution as follows: we interpret a substitution \(\Gamma \stnt \sigma: \Delta\) as a function \(\dnt{\Gamma \stnt \sigma: \Delta}: \dnt{\Delta} \to \dnt{\Gamma}\) by composing it elementwise with the the denotational semantics. Supposing \(D \in \dnt{\Delta}\):
\begin{equation}
    \pi_{x, \Gamma}(\dnt{\Gamma \stnt \sigma: \Delta}\;D) = \dnt{\Delta \stnt \sigma\;x: A}\;D \in \dnt{A}
\end{equation}
We may now state semantic substitution for the \stlc{} as follows: 
\begin{lemma}[Semantic Substitution (\stlc)]
    Given \stlc{} derivation \(\Gamma \stnt a: A\) and \stlc{} substitution \(\Gamma \stnt \sigma: \Delta\), we have
    \begin{equation*}
        \dnt{\Gamma \stnt a: A} \circ \dnt{\Gamma \stnt \sigma: \Delta} = \dnt{\Delta \stnt [\sigma]a: [\sigma]A}
    \end{equation*}
    \label{lem:subst-denot-commute}

    \formalization{LogicalRefinement/Stlc/Subst.lean}{theorem Stlc.HasType.subst_interp_dist}
\end{lemma}

\begin{figure}
    \begin{gather*}
        \prftree{}{\Gamma \stnt (): \mb{1}} \qquad
        \prftree{}{\Gamma \stnt \ms{error}: A} \qquad
        \prftree
            {\Gamma \stnt a: A}
            {\Gamma, x: A \stnt e: B}
            {\Gamma \stnt \ms{let}\;x = a\;\ms{in}\;e: B}
        \\
        \prftree{\Gamma, x: A \stnt e: B}
            {\Gamma \stnt \lambda x: A, e: A \to B} \qquad
        \prftree{\Gamma \stnt f: A \to B}{\Gamma \stnt a: A}
            {\Gamma \stnt f\ a: B} \qquad
        \prftree{\Gamma \stnt l: A}{\Gamma \stnt r: B}
            {\Gamma \stnt (l, r): A \times B} \\
        \prftree{\Gamma \stnt e: A}{\Gamma \stnt \ms{inl}\;e: A + B} \qquad
        \prftree{\Gamma \stnt e: B}{\Gamma \stnt \ms{inr}\;e: A + B} \qquad
        \prftree
            {\Gamma \stnt e: A + B}
            {\Gamma, x: A \stnt l: C}
            {\Gamma, y: B \stnt r: C}
            {\Gamma \stnt \ms{cases}\;e
                \;(\ms{inl}\;x \mapsto l)
                \;(\ms{inr}\;y \mapsto r): C} \\
        \prftree{}{\Gamma \stnt 0: \nats} \qquad
        \prftree{}{\Gamma \stnt \ms{succ}: \nats \to \nats} \qquad
        \prftree
            {\Gamma \stnt e: \nats}
            {\Gamma \stnt z: C}
            {\Gamma, x: C \stnt s: C}
            {\Gamma \stnt \ms{natrec}\;e\;z\;(x \mapsto s): C}
    \end{gather*}
    \caption{\stlc{} typing rules}
    \Description{
        STLC typing rules
    }
    \label{fig:stlc-typing}
\end{figure}

\begin{figure}
    \begin{gather*}
        \dnt{\mb{0}} = \{\}, \qquad 
        \dnt{\mb{1}} = \{*\}, \qquad
        \dnt{\nats} = \nats \\
        \dnt{A \to B} = \dnt{A} \to \ms{M}\dnt{B}, \qquad
        \dnt{A + B} = \dnt{A} \sqcup \dnt{B}, \qquad
        \dnt{A \times B} = \dnt{A} \times \dnt{B}
    \end{gather*}
    \caption{\stlc{} type denotations parameterized by a monad \(\ms{M}\). The denotation of a term of type \(A\) has type \(\ms{M}\dnt{A}\).}
    \Description{
        STLC type denotations parameterized by the exception monad M
    }
    \label{fig:stlc-ty-denot}
\end{figure}

\begin{figure}
    \begin{equation*}
        \boxed{\dnt{\Gamma \stnt a: A}: \dnt{\Gamma} \to \ms{M}\;\dnt{A}}
    \end{equation*}
    \begin{align*}
        \dnt{\Gamma \stnt x: A}\;G ={}& \pi_{x, \Gamma}G \\
        \dnt{\Gamma \stnt (): \mb{1}}\;G ={}& \ms{ret}\;() \\
        \dnt{\Gamma \stnt \ms{error}: A}\;G ={}& \ms{error}_A \\
        \dnt{\Gamma \stnt \lambda x: A, e: A \to B}\;G ={}&
        \ms{ret}\;(\lambda a, \dnt{\Gamma, x: A \stnt e: B}(G, \ms{ret}\;a)) \\
        \dnt{\Gamma \stnt f\;a: B}\;G ={}& 
        \ms{bind}\;(\dnt{\Gamma \stnt f: (x: A) \to B}\;G) 
            \;(\lambda f, \ms{bind}\;(\dnt{\Gamma \stnt a: A}\;G)\;f) \\
        \dnt{\Gamma \stnt (l, r): A \times B}\;G
        ={}& \ms{bind}\;(\dnt{\Gamma \stnt l: A}\;G) 
        \;(\lambda l, \ms{bind}\;(\dnt{\Gamma \stnt r: B}\;G)
        \;(\lambda r, \ms{ret}\;(l, r)))\\
        \dnt{\Gamma \stnt \ms{inl}\;e: A + B}\;G
        ={}& \ms{fmap}\;\ms{inl}\;(\dnt{\Gamma \stnt e: A}\;G) \\
        \dnt{\Gamma \stnt \ms{inr}\;e: A + B}\;G
        ={}& \ms{fmap}\;\ms{inr}\;(\dnt{\Gamma \stnt e: B}\;G) \\
        %
        %TODO: adjust formatting here...
        \ednt{
        \arraycolsep=1pt    
        \begin{array}{ll} 
            \Gamma \stnt& \ms{cases}\;d \\
            &\quad(\ms{inl}\;y \mapsto l) \\ 
            &\quad(\ms{inr}\;z \mapsto r) : C \end{array}} \;G 
        ={}& \left({
            \arraycolsep=0pt
            \begin{array}{l}
                \ms{bind}\;(\dnt{\Gamma \stnt d: A + B}\;G)\;(\lambda d, \\
                \quad \ms{cases}\;d \\
                \qquad(\ms{inl}\;a \mapsto 
                    \dnt{\Gamma, y: A \stnt l: C}\;(G, \ms{ret}\;a)) \\
                \qquad(\ms{inr}\;b \mapsto 
                    \dnt{\Gamma, z: B \stnt r: C}\;(G, \ms{ret}\;b))
            \end{array}
        }\right) \\
        \dnt{0}\;G ={}& \ms{ret}\;0 \\
        \dnt{\ms{succ}}\;G ={}& \ms{ret}\;\ms{succ} \\
        \dnt{\Gamma \stnt \ms{natrec}\;n\;z\;(x \mapsto s) : C} \;G 
        ={}& \left({
            \arraycolsep=0pt
            \begin{array}{l}
                \ms{bind}\;(\dnt{\Gamma \stnt n: \nats}\;G)\;(\lambda n, \\
                    \quad \ms{natrec}\;n\;(\dnt{\Gamma \stnt z: C}\;G) \\
                        \qquad (c \mapsto \ms{bind}\;c\;(\lambda c.
                            \dnt{\Gamma, x: C \vdash s: C}(G, \ms{ret}\;c)))
            \end{array}
        }\right)
        \\
        \dnt{\Gamma \stnt \ms{let}\;x = a\;\ms{in}\;e: B}\;G
        ={}&
        \ms{bind}\;
        (\dnt{\Gamma \stnt a: A}\;G)
            \;(\lambda a', \dnt{\Gamma, x: A \stnt e: B}(G, \ms{ret}\;a'))
    \end{align*}
    \caption{
        Denotations for \stlc{} terms, where \(\ms{M}\) is the exception monad with
        \(\ms{error}_A: \ms{M}\;A\)
    }
    \Description{Denotations for STLC terms}
    \label{fig:stlc-term-denot}
\end{figure}

\subsection{Erasure}
\label{ssec:erasure}
We define a notion of erasure \(|A|\) of \ert{} types and terms to corresponding \stlc{} ones in figure \ref{fig:stlc-erasure}. Erasure on types simply erases all dependency and propositional information leaving behind a simply-typed skeleton. For tuple-like type formers like \(\lrset{x}{A}{\varphi}\), the propositional information is erased completely, yielding \(|A|\), whereas for function-like type formers like \(\lrasm{u}{\varphi}{A}\), it is instead erased to a unit, yielding \(\mb{1} \to |A|\); this is to avoid issues with eager evaluation.
Where necessary, we take proofs and propositions to erase into the unit as a convenience, i.e.,
\begin{equation}
    \forall \varphi \in \ms{Prop}, \ers{\varphi} = \mb{1} \in \tystlc,
    \qquad
    \forall p \in \ms{Proof}, \ers{p} = () \in \tmstlc
\end{equation}
We may then recursively define the erasure of an \ert{} context into an \stlc{} context as follows:
\begin{equation}
    \ers{\cdot} = \cdot, \qquad
    \ers{\Gamma, x: A} = \ers{\Gamma}, x: \ers{A} \qquad
    \ers{\Gamma, u: \varphi} = \ers{\Gamma}, u: \mb{1} \qquad
    \ers{\Gamma, \|x: A\|} = \ers{\Gamma}, x: \mb{1}
\end{equation}
As one would expect; erasing a well-typed \ert{} term yields a well-typed \stlc{} term; in particular, we have that:
\begin{lemma}[Erasure]
    Given a derivation \(\Gamma \lrnt a: A\), we may derive a derivation of \(\ers{\Gamma} \stnt \ers{a}: \ers{A}\), written \(\ers{\Gamma \lrnt a: A}\)
    \label{lem:erase}

    \formalization{LogicalRefinement/Stlc/Interp.lean}{theorem HasType.stlc}
\end{lemma}
With this definition in hand, we may define the erasure of a substitution pointwise, that is, as
\begin{equation}
    \forall \sigma \in \ms{Var} \to \ms{Term}, \ers{\sigma}\;a = \ers{\sigma\;a}
\end{equation}
It then follows as a trivial corollary of lemma~\ref{lem:erase} that, given a substitution \(\Gamma \lrnt \sigma: \Delta\), we have \(\ers{\Gamma} \stnt \ers{\sigma}: \ers{\Delta}\); we write this as \(\ers{\Gamma \lrnt \sigma: \Delta}\).
We are now in a position to prove that the erasure of the substitution of an \ert{} term is the erased substitution of the corresponding erased \stlc{} term:
\begin{lemma}[Substitution and Erasure Commute]
    Given a substitution \(\Gamma \lrnt \sigma: \Delta\), we have \(\ers{[\sigma]a} = [\ers{\sigma}]\ers{a}\)
    \label{lem:erase-comm}

    \formalization{LogicalRefinement/Stlc/InterpSubst.lean}{theorem HasType.subst_stlc_commute}
\end{lemma}
We may then deduce the following corollary from lemma~\ref{lem:subst-denot-commute} and lemma~\ref{lem:erase-comm}: 
\begin{corollary}[Substitution and Erasure Denotation Commute] 
    Given \ert{} derivation \(\Gamma \lrnt a: A\) and \ert{} substitution \(\Gamma \lrnt \sigma: \Delta\), we have
    \begin{equation*}
        \dnt{\ers{\Gamma \lrnt a: A}} \circ \dnt{\ers{\Gamma \lrnt \sigma: \Delta}} =
        \dnt{\ers{\Delta \lrnt [\sigma]a: [\sigma]A}}
    \end{equation*}
    \label{lem:subst-term-commute}

    \formalization{LogicalRefinement/Stlc/InterpSubst.lean}{theorem HasType.subst_stlc_commute}
\end{corollary}

\begin{figure}
    \begin{equation*}
        \boxed{\ers{\cdot}: \ms{Type} \to \tystlc}
    \end{equation*}
    \begin{gather*}
        \ers{(x: A) \to B} = \ers{A} \to \ers{B}, \qquad
        \ers{(x: A) \times B} = \ers{A} \times \ers{B} \\
        \ers{\lrimp{u}{\varphi}{A}} = \mb{1} \to \ers{A}, \qquad
        \ers{\lrset{x}{A}{\varphi}} = \ers{A}, \qquad
        \ers{\forall x: A, B} = \mb{1} \to \ers{B}, \qquad
        \ers{\exists x: A, B} = \ers{B} \\
        \ers{\mb{1}} = \mb{1}, \qquad
        \ers{A + B} = \ers{A} + \ers{B}, \qquad 
        \ers{\nats} = \nats \\
    \end{gather*}
    \begin{equation*}
        \boxed{\ers{\cdot}: \ms{Term} \to \tmstlc}
    \end{equation*}
    \begin{gather*}
        \ers{x} = x \qquad
        \ers{\lambda x: A, e} = \lambda x: \ers{A}. \ers{e} \qquad
        \ers{a\;b} = \ers{a}\;\ers{b}, \qquad
        \ers{(a, b)} = (\ers{a}, \ers{b}) \\
        \ers{\ms{let}\;(x, y): A = e\;\ms{in}\;e'} =
        \ms{let}\;(x, y) = \ers{e}\;\ms{in}\;\ers{e'}, \qquad
        \ers{\ms{inl}\;e} = \ms{inl}\;\ers{e}, \qquad
        \ers{\ms{inr}\;e} = \ms{inr}\;\ers{e} \\
        \ers{\ms{cases}[x \mapsto C]\;e
            \;(\ms{inl}\;y \mapsto l)
            \;(\ms{inr}\;z \mapsto r)} =
        \ers{\ms{cases}\;e
            \;(\ms{inl}\;y \mapsto \ers{l})
            \;(\ms{inr}\;z \mapsto \ers{r})} \\
        \ers{\lambda u: \varphi, e} = \lambda\_:\mb{1}.\ers{e}, \qquad
        \ers{a\;p} = \ers{a}\;(), \qquad
        \ers{\{a, p\}} = \ers{a} \\
        \ers{\ms{let}\;\{x, y\}: A = e\;\ms{in}\;e'} =
        \ms{let}\;x = \ers{e}\;\ms{in}\;\ers{e'} \\
        \ers{\lambda\|x: A\|, e} = \lambda\_:\mb{1}.\ers{e}, \qquad
        \ers{a\;\|b\|} = \ers{a}\;(), \qquad
        \ers{(\|a\|, b)} = \ers{b} \\
        \ers{\ms{let}\;(\|x\|, y): A = e\;\ms{in}\;e'} =
        \ms{let}\;y = \ers{e}\;\ms{in}\;\ers{e'} \\
        \ers{0} = 0 \qquad 
        \ers{\ms{succ}} = \ms{succ} \qquad
        \ers{\ms{natrec}[x \mapsto C]\;e\;z\;(\|\ms{succ}\;n\|, y \mapsto b)}
        = \ms{natrec}\;\ers{e}\;\ers{z}\;(y \mapsto \ers{b}) \\
        % \ers{\ms{subst}[x \mapsto B][a][b]\;p\;e} =
        % \ers{e}, \qquad
        \ers{\ms{absurd}\;p} = \ms{error}
    \end{gather*}
    \caption{Erasure of \ert{} to \stlc{}}
    \Description{Erasure of ERT to STLC}
    \label{fig:stlc-erasure}
\end{figure}

\subsection{Denotational Semantics}
\label{ssec:semantics}

Using lemma~\ref{lem:erase}, we could assign a computational meaning to well-typed \ert{} terms \(\Gamma \vdash t: A\) by simply composing the denotation for \stlc{} terms with the erasure function, i.e., taking \(\dnt{\ers{\Gamma \vdash t: A}}: \dnt{\ers{\Gamma}} \to \ms{M}\dnt{\ers{A}}\). While this interpretation assigns terms a computational meaning, it simply ignores their refinements: there is not yet any guarantee that the refinements mean anything.

To rectify this, we will give a denotational semantics for \ert{} types (in figure~\ref{fig:ert-denot}), which maps each type to a subset of the denotation of the corresponding erased type. This semantics is mutually recursive with semantics for \ert{} propositions as well.

The denotation of types (and propositions) is parameterized by an environment \(G\) drawn from the interpretation of the context \(\Gamma\). To break the recursion between the semantics of contexts and types, the domain of the interpretation function for types (and propositions) is not restricted to valid \ert{} environments \(G \in \dnt{\Gamma\;\ms{ok}}\), but is defined for all \stlc{} environments \(G \in \dnt{\ers{\Gamma}}\). However, the semantics of \ert{} types does depend upon the values of ghost variables. As a result, we do not consider the erasure \(\ers{\Gamma}\), but rather the erasure of the \emph{upgrade} \(\ers{\upg{\Gamma}}\).

The denotation of types basically follows the structure of a unary logical relation (i.e., a logical predicate). For example, the intepretation \(\dnt{\Gamma \vdash (x:A) \to B}\;G\) are functions \(\ers{A} \to \ms{M}(\ers{B})\) which satisfy the property that for all inputs in the refined type \(A\), the function returns a pure value in the refined type \(B\). An element of a pair \((x:A) \times B[x]\) is a pair \((a, b)\) in \(\ers{A} \times \ers{B}\) where \(a\) is an element of \(A\) and \(b\) is an element of \(B[a]\). 

The union type \(\exists x:A, B[x]\) are those elements of \(|B|\) which lie in \(B[a]\) for some \(a\) in \(A\). The intersection type \(\forall x:A, B\) is almost dual, but to account for call-by-value evaluation in the case where \(A\) may be an empty type, we consider thunks \(1 \to \ms{M}(\ers{B})\) rather than elements of \(\ers{B}\). This is also where the terms "intersection type" and "union type" come from -- the semantics of \(\forall x:A, B\) is literally a giant intersection, and likewise \(\exists x:A, B[x]\) is interpreted with a giant union. 

An element of a subset type \(\{x:A \mid \varphi\}\) is an element of
\(A\) which also satisfies the property \(\varphi\). The dual
precondition type \(\lrasm{u}{\varphi}{A}\) represents elements of \(A\),
conditional on \(\phi\) holding. Just as with intersections, this type
must be represented by thunks \(1 \to \ms{M}(\ers{A})\) to account for
the case where \(\varphi\) is false. Units and natural numbers have
the same denotation as their simply-typed counterpart, and a coproduct
\(A + B\) is either a left injection of a value satisfying \(A\) or a
right injection of a value satisfying \(B\).  A proposition
\(\varphi\) could be interpreted by a map
\(\dnt{\ers{\Gamma}} \to \mb{2}\), but it is more convenient to think
of it as a subset of \(\dnt{\ers{\Gamma}}\) -- the set of contexts for
which the proposition holds. So \(\top\) is the whole set of contexts,
\(\bot\) is the empty set, and disjunction and conjunction are
modelled by union and intersection. Because conjunction is written
\(u:\varphi \land \psi[u]\), we have to extend the environment of the
interpretation of \(\psi\). Since we erase all propositions to
\(\mb{1}\), we just choose \(\ms{ret}\;()\) as the erased proof. The
same idea is used in the case of propositional
implication. Quantifiers in our language of propositions are
interpreted by quantifiers in the meta-language, and equality is
interpreted as the set of contexts \(G \in \dnt{\ers{\upg{\Gamma}}}\)
for which the equality holds. Note in particular that this means that
the law of the excluded middle is sound, and cannot interfere with
program execution in any way (since all propositions are erased).

Because the unrefined language is ambiently effectful (with the effect
of error stops), every expression lies in a monadic type in the
semantics. The refined semantics of the monad is given by the
$\mc{E}\dnt{\Gamma \lrnt A\;\ms{ty}}$ relation, which picks out the
subset of the monad not equal to $\ms{error}_A$. That is, we require
all refined terms to terminate without error.

Propositions are interpreted as a map from the intepretation of
contexts into the (boolean, classical) truth values, or equivalently,
as a subset of the contexts. Hence each logical connective is
interpreted by the corresponding operation in the Boolean algebra of
sets -- conjunction is intersection, disjunction is union, and so
on. We interpret equality at a type $A$ as equality of the underlying
elements of $|A|$.

This interpretation, while simple, is in some sense
\emph{too} simple -- it causes function extensionality to fail.  In
particular, suppose $\ms{Pos} \equiv \{ x:\ints \mid x > 0 \}$ is the
type of positive numbers. Two functions $f, g : \ms{Pos} \to \ints$
may agree on all positive arguments but fail to be equal because they
erase to two functions which do different things on negative
arguments. (For example, consider $f = \ms{id}$ and $g = \ms{abs}$.)
This is also the reason that the \brle{$\eta_{\ms{ir}}$} and
\brle{$\eta_{\ms{pr}}$} rules have an inhabitation premise (e.g.,
$\upg{\Gamma} \lrnt i : A$). A more complex semantics based on partial
equivalence relations could let us resolve this issue, but we
preferred to stick with the simplest possible semantics for expository
reasons.

The semantics of types and propositions is defined over a bigger set of contexts than just the valid ones, but once we have this semantics, we can use it to define the valid contexts. Again, the interpretation \(\dnt{\Gamma\;\ms{ok}}\) is going to be the subset of \(\dnt{\ers{\upg{\Gamma}}}\) which satisfy all the propositions and in which all the values lie within their \ert{} types. So the empty context is inhabited by the empty environment; the context \(\Gamma, u:\varphi\) is inhabited by \((G, \ms{ret}\;())\) when \(G\) is in \(\dnt{\Gamma\;\ms{ok}}\) and \(\varphi\) is satisfied; and \(\Gamma, x:A\) is inhabited by \((G, \ms{ret}\;x)\) when \(G \in \dnt{\Gamma\;\ms{ok}}\) and \(x\) is in the (subset) interpretation of \(A\). (The case of ghost variables is the same as ordinary variables, since we care about the values of ghost variables when interpreting propositions in refined types.) 
It is easy to show that no \ert{} type can contain any errors. 
\begin{lemma}[Convergence]
    Given an \ert{} derivation \(\Gamma \lrnt A\;\ms{ty}\), we have
    \(\forall G, \ms{error} \notin \dnt{\Gamma \lrnt A\;\ms{ty}}\;G\)
    \label{lem:term}

    \formalization{LogicalRefinement/Denot/Basic.lean}{theorem HasType.denote_ty_non_null}
\end{lemma}
However, we give semantics to \ert{} \emph{terms} by erasure, and so we have to connect the erased semantics of \ert{} terms to these semantic types. Furthermore, the semantics of \ert{} types cares about ghost values, but the semantics of erased terms ignore ghost values.

To relate these two, we also need to define the corresponding notion of a downgrade
of an environment. Given an environment \(G \in \dnt{\ers{\upg{\Gamma}}}\), we recursively define its downgrade \(\dng{G}_\Gamma \in \dnt{\ers{\Gamma}}\), written \(\dng{G}\) when \(\Gamma\) is clear from context, as
\begin{equation}
    \dng{*}_{\cdot} = *, \qquad
    \dng{(x, G)}_{u: \varphi, \Gamma} = (x, \dng{G}_\Gamma), \qquad
    \dng{(y, G)}_{x: A, \Gamma} = (y, \dng{G}_\Gamma), \qquad
    \dng{(y, G)}_{\|x: A\|, \Gamma} = (*, \dng{G}_\Gamma)
\end{equation}
Semantically, this discards all of the ghost information from an environment.
We may now state the primary theorems proven about the semantics of \ert{}, namely, semantic substitution and semantic regularity. 
\begin{theorem}[Semantic Substitution]
    Given \ert{} derivation \(\Gamma \lrnt A\;\ms{ty}\),
    \ert{} substitution \(\Gamma \lrnt \sigma: \Delta\), 
    and valid environment \(D \in \dnt{\Delta\;\ms{ok}}\), we have
    \begin{equation*}
        \dnt{\Delta \lrnt [\sigma]A\;\ms{ty}}\;D =
        \dnt{\Gamma \lrnt A\;\ms{ty}}(\dnt{\ers{\upg{\Gamma} \lrnt \upg{\sigma}: \upg{\Delta}}}\;D)
    \end{equation*}
    \label{thm:semsub}
    \formalization{LogicalRefinement/Denot/Subst.lean}{theorem SubstCtx.subst_denot}
  \end{theorem}
% \begin{proof}
%     By induction over \ert{} derivations. For details, see \tty{theorem SubstCtx.subst_denot} in formalization source file \tty{LogicalRefinement/Denot/Subst.lean}
% \end{proof}
We can also show that for any well-typed \ert{} term, its erasure lies in the interpretation of the \ert{} type. This shows that every well-typed term satisfies the properties of its fancy type. 
\begin{theorem}[Semantic Regularity] 
    Given an \ert{} derivation \(\Gamma \lrnt a: A\), we have
    \begin{equation*}
        \forall G \in \dnt{\Gamma\;\ms{ok}}, 
            \dnt{\ers{\Gamma \lrnt a: A}}\;\dng{G} \in 
            \dnt{\Gamma \lrnt A\;\ms{ty}}\;G
    \end{equation*}
    \label{thm:sound}
    \formalization{LogicalRefinement/Denot/Regular.lean}{theorem HasType.denote}
\end{theorem}
% For both of these propositions, similar theorems hold for contexts and propositions; see the appendix for details. 
  
\TODO{check this} 

Furthermore, it is an immediate corollary of lemma~\ref{lem:term} and theorem~\ref{thm:sound} that the \ert{} logic is consistent, since $\left\{x:1 \mid \phi \right\}$ is inhabited if and only if $\phi$ is true. It is also a corollary that for any well-typed term \(\Gamma \lrnt a: A\), we have that
\(
    \forall G \in \dnt{\Gamma\;\ms{ok}}, \dnt{\ers{\Gamma \lrnt a: A}}\;G \neq \ms{error}
\).
That is, "well-typed programs do not go wrong".

\begin{figure}
    \begin{equation*}
        \boxed{\dnt{\Gamma \lrnt A\;\ms{ty}}: 
            \dnt{\ers{\upg{\Gamma}}} \to \mc{P}(\dnt{\ers{A}})}
    \end{equation*}
    \begin{align*}
        \dnt{\Gamma \lrnt \mb{1}\;\ms{ty}}\;G ={}& \{()\} \\
        \dnt{\Gamma \lrnt (x: A) \to B\;\ms{ty}}\;G ={}& 
            \{f \in \dnt{\ers{A}} \to \ms{M}\dnt{\ers{B}}
                \mid
                    \forall x \in \dnt{\Gamma \lrnt A\;\ms{ty}}\;G, 
                    \\&\qquad 
                    f\;x \in 
                    \mc{E}\dnt{\Gamma, x: A \lrnt B\;\ms{ty}}\;(G,\ms{ret}\;x)
                    \} \\
        \dnt{\Gamma \lrnt (x: A) \times B\;\ms{ty}}\;G ={}&
             \{(l, r) 
             \mid l \in \dnt{\Gamma \lrnt A\;\ms{ty}}\;G 
             \land r \in \dnt{\Gamma, x: A \lrnt B\;\ms{ty}}\;(G, \ms{ret}\;l) \}\\
        \dnt{\Gamma \lrnt A + B\;\ms{ty}}\;G ={}& 
            \ms{inl}(\dnt{\Gamma \lrnt A\;\ms{ty}}\;G) 
            \cup \ms{inr}(\dnt{\Gamma \lrnt B\;\ms{ty}}\;G) \\
        \dnt{\Gamma \lrnt \lrasm{u}{\varphi}{A}\;\ms{ty}}\;G ={}&
            \{
                f \in \mb{1} \to \ms{M}\dnt{\ers{A}} \mid
                \\&\qquad G \in 
                    \dnt{\Gamma \vdash \varphi\;\ms{pr}}
                    \implies f\;() \in 
                    \mc{E}\dnt{\Gamma, u: \varphi \vdash A\;\ms{ty}}\;(G, \ms{ret}\;())
            \} \\
        \dnt{\Gamma \lrnt \lrset{x}{A}{\varphi}\;\ms{ty}}\;G ={}&
            \{a \in \dnt{\Gamma \lrnt A\;\ms{ty}}\;G
                \mid (G, a) \in \dnt{\Gamma, x: A \lrnt \varphi\;\ms{pr}}\} \\
        \dnt{\Gamma \lrnt \forall x: A, B\;\ms{ty}}\;G ={}&
            \{
                f \in \mb{1} \to \ms{M}\dnt{\ers{B}} \mid
                \\&\qquad
                    \forall x \in \dnt{\Gamma \lrnt A\;\ms{ty}}\;G,
                        f\;() \in \mc{E}\dnt{\Gamma, x: A \lrnt B\;\ms{ty}}
                            \;(G, \ms{ret}\;x)
            \} \\
        \dnt{\Gamma \lrnt \exists x: A, B\;\ms{ty}}\;G ={}&
            \bigcup_{x \in \dnt{\Gamma \lrnt A\;\ms{ty}}\;G}
                \dnt{\Gamma, x: A \lrnt B\;\ms{ty}}(G, \ms{ret}\;x) \\
        \dnt{\Gamma \lrnt \nats\;\ms{ty}}\;G ={}& \nats \\
        \mc{E}\dnt{\Gamma \lrnt A\;\ms{ty}} 
        ={}& \lambda G, \ms{ret}\;(\dnt{\Gamma \stnt A\;\ms{ty}}\;G):
        \dnt{\ers{\upg{\Gamma}}} \to \mc{P}(\ms{M}\dnt{\ers{A}})
    \end{align*}
    \begin{equation*}
        \boxed{\dnt{\Gamma \lrnt \varphi\;\ms{pr}}: 
            \mc{P}(\dnt{\ers{\upg{\Gamma}}})}
    \end{equation*}
    \begin{align*}
        \dnt{\Gamma \lrnt \top\;\ms{pr}} ={}& \dnt{\ers{\upg{\Gamma}}} \\
        \dnt{\Gamma \lrnt \bot\;\ms{pr}} ={}& \varnothing \\
        \dnt{\Gamma \lrnt \lrimp{u}{\varphi}{\psi}\;\ms{pr}} ={}&
            \{G \mid 
                G \in \dnt{\Gamma \lrnt \varphi\;\ms{pr}}
                \implies (G, \ms{ret}\;()) \in \dnt{\Gamma, u: \varphi \lrnt \psi\;\ms{pr}
            }\} \\
        \dnt{\Gamma \lrnt (u: \varphi) \land \psi\;\ms{pr}} ={}&
            \{G \mid 
                G \in \dnt{\Gamma \lrnt \varphi\;\ms{pr}}
                \land (G, \ms{ret}\;()) \in \dnt{\Gamma, u: \varphi \lrnt \psi\;\ms{pr}
            }\} \\
        \dnt{\Gamma \lrnt \varphi \lor \psi\;\ms{pr}} ={}&
            \dnt{\Gamma \lrnt \varphi\;\ms{pr}} 
            \cup \dnt{\Gamma \lrnt \psi\;\ms{pr}} \\
        \dnt{\Gamma \lrnt \forall x: A, \varphi\;\ms{pr}} ={}&
            \{G \mid
                \forall x \in \dnt{\Gamma \lrnt A\;\ms{ty}}\;G,
                (G, \ms{ret}\;x) \in \dnt{\Gamma, x: A \lrnt \varphi\;\ms{pr}}
            \} \\
        \dnt{\Gamma \lrnt \exists x: A, \varphi\;\ms{pr}} ={}&
            \{G \mid
                \exists x \in \dnt{\Gamma \lrnt A\;\ms{ty}}\;G,
                (G, \ms{ret}\;x) \in \dnt{\Gamma, x: A \lrnt \varphi\;\ms{pr}}
            \} \\
        \dnt{\Gamma \lrnt a =_A b\;\ms{pr}} ={}&
            \{G \mid 
                \dnt{\ers{\upg{\Gamma} \lrnt a: A}}\;G =
                \dnt{\ers{\upg{\Gamma} \lrnt b: A}}\;G
            \}
    \end{align*}
    \begin{equation*}
        \boxed{\dnt{\Gamma\;\ms{ok}}: \mc{P}(\dnt{\ers{\upg{\Gamma}}})}
    \end{equation*}
    \begin{align*}
        \dnt{\cdot\;\ms{ok}} ={}& \{\cdot\} \\
        \dnt{\Gamma, x: A\;\ms{ok}} 
        = \dnt{\Gamma, \|x: A\|\;\ms{ok}} ={}& \{
            (G, \ms{ret}\;x) \mid 
                G \in \dnt{\Gamma\;\ms{ok}} \land 
                x \in \dnt{\Gamma \lrnt A\;\ms{ty}}\;G
        \} \\
        \dnt{\Gamma, u: \varphi\;\ms{ok}} ={}&  
            (\dnt{\Gamma\;\ms{ok}} \cap \dnt{\Gamma \lrnt \varphi\;\ms{pr}})
            \times \ms{M}\mb{1}
    \end{align*}
    \caption{Denotations for \ert{}}
    \Description{Denotations for \ert{}}
    \label{fig:ert-denot}
\end{figure}

\section{Formal Verification}

\label{sec:lean4}

We have proved all the results stated in the previous sections in Lean 4. The proof development is about 15.8 kLoC in length and is partially automated, though there is much potential for further automation. In particular, lemmas \ref{lem:upgrade}, \ref{lem:downgrade}, and \ref{lem:subst} are heavily automated, with one template tactic that applies to most of the cases,  whereas theorems \ref{thm:semsub} and \ref{thm:sound} and lemmas \ref{lem:subst-denot-commute}, \ref{lem:erase-comm}, and \ref{lem:subst-term-commute} had to be proved manually. These properties typically had lots of equality coercions, which had to be plumbed manually because they are difficult to automate. It is possible that with more experience with Lean tactics they might also be automated, but we were not able to do so. 

The formalized syntax and semantics are mostly the same as that presented in this writeup. There are two main differences. First, we have implemented variables using de-Bruijn indices. Second, we folded types, propositions, terms, and proofs into a single inductive type to avoid mutual recursion, which Lean 4 currently has poor support for.

This project was the first time the authors used Lean 4 for serious formalization work. While we ran into numerous issues due to Lean still being in active early-stage development, we found it to be a highly effective formalization tool. One issue we ran into was very high memory usage and, in some cases, timeouts, when using Lean's \tty{simp} tactic on complex pattern matches. The addition of the \tty{dsimp} tactic, after a discussion on the Lean 4 Zulip, mostly alleviated this, and performance has improved in later versions of Lean. We otherwise found the quality of automation to be very good: even though the authors are novices at Lean, we were able to easily maintain and extend the proofs without needing to edit theorems proved via tactics. For example, we originally forgot to include the \hyperref[fig:ert-axiom-rule]{\brle{Unit-WF}} axiom, but we were able to include it with only minor edits to the manual theorems in about 30 minutes. As the formalization made heavy use of dependent types, we also ran into many issues attempting to establish equalities between dependently typed terms. However, in this case, we found Lean relatively easy to use compared to other dependently-typed proof assistants based on dependent types, with our experiments at Coq-based formalization running into similar issues.

\section{Discussion}

\label{sec:discussion}

\paragraph{Function Extensionality, Recursion and Effects}
  Our current semantics is inconsistent with function extensionality because two functions must be equal over their entire, unrefined domain to satisfy the denotation of the equality type. To support extensionality (and related types like quotient types), we should be able to intepret the calculus via a semantics based on partial equivalence relations, as in \cite{per-operational}. 

We also want to reason about the partial correctness and divergent programs. Hence, it makes sense to add support for general recursive definitions, including nonterminating definitions, by moving to a domain-theoretic semantics (rather than the set-theoretic semantics we currently use). 
We also want to extend the base language with more effects (such as store and IO) and extend \ert{} to support fine-grained reasoning about them via an effect system such as in \cite{pems}. 

\paragraph{Categorical Semantics}
The motivating model of refinement types underlying our work is that of \cite{ftrs}, which equates type refinement systems with functors from a category of typing derivations to a category of terms. In essence, one can view our work as taking the setup in \cite{ftrs} and inlining all the categorical definitions for the case of the simply-typed lambda calculus.

We would like to update our formalisation to work in terms of the categorical semantics; this would let us to account for all of the extensions above at once, without having to reprove theorems (such as semantics substitution and regularity) for each modification. 

Recently, \citet{Kura21} has studied the denotational semantics of a
system of refinements over Ahman's variant of dependent
call-by-push-value~\cite{ahman}. Kura also uses a fibrational
semantics similar to the Zeilberger-Mellies semantics (as well as
Katsumata's semantics of effect systems~\cite{pems}) to equip a
dependent type theory with a subtype relation arising from the
entailment relation of first-order logic. Unlike \ert{}, the use of
subtyping means that there are no explicit proofs, and hence type
checking is undecidable. 

\paragraph{Units in the Erasure}
\label{par:units-in-the-erasure}

The erasure function for \ert introduces many units as part of the translation (e.g., $|u:\phi \Rightarrow A| = \mb{1} \to |A|$).
This is necessary to ensure the property that no well-typed term can signal an error. Consider the example:
\begin{displaymath}
    \hat{\lambda}||u:\bot||. \ms{absurd}(u)  : (u:\bot) \Rightarrow A
\end{displaymath}
This erases to the term:
\begin{displaymath}
  \lambda x:\mb{1}. \ms{error} 
\end{displaymath}
Without the $\lambda x:\mb{1}$, the term would erase to $\ms{error}$, which violates our invariant.

It is possible to avoid introducing any units at all if the base
calculus is in call-by-push-value form. However, since one of our main
objectives was to have a simple, familiar semantics, we the price of
introducing units was lower than introducing call-by-push-value.

\paragraph{Automation and Solver Integration}

One of the critical advantages of refinement types is the potential for significantly reducing the annotation burden of formal verification. Hence, to make \ert{} usable, it should be able to be automated to a similar degree for similarly complex programs. One potential form of basic automation is support for an "\(\ms{smt}\)" tactic, similar to section~\ref{sec:examples}'s "\(\beta\)" tactic; we can similarly envision calling out to various automated theorem provers like Vampire \cite{vampire} or SPASS \cite{spass}.

A more powerful approach would be to adapt the work on Liquid Typing \cite{liquid-types} to this setting, which works by inferring appropriate refinement types and proofs for an unrefined program such that, given that the program's preconditions are satisfied, the preconditions of all function calls within the program as well as the postconditions of the program are both satisfied. Liquid Typing sometimes requires annotations to infer appropriate invariants and may require explicit checks to be added for conditions it cannot verify are implied by the preconditions. One way to combine Liquid Typing with \ert{} would be to, using \ert{} types as annotations, automatically refine the types of subterms of an \ert{} program to make it typecheck, inferring and inserting proofs as necessary. One advantage of this approach would be that (assuming it compiles down to fully-annotated \ert{}) it removes the liquid typing algorithm itself from the trusted codebase, and, if the solvers used support proof output, the solvers themselves as well. Furthermore, we could replace potentially expensive runtime checks with explicit proofs.

\section{Related Work}
\label{sec:related-work}

\subsection{Relationship of \texorpdfstring{\ert{}}{LambdaERT} to Dependent Types}

The most well-known approach to designing programming languages with
integrated support for proof is dependent type theory. The semantics
of dependent type theories is generally given in an "intrinsic"
style, in which well-typed terms (in fact, typing derivations) are
given a denotational semantics, and ill-typed terms are regarded as
meaningless (i.e., have no semantics).

On the other hand, \ert{} is a refinement type system, which takes an
existing programming language (in this case, the simply-typed lambda
calculus with error stops), and extends it so that existing programs
can be given richer, more precise types. This ensures that it is
always possible to forget the rich types and be left with the
simply-typed skeleton. A good analogy is to Hoare logic, in which
pre-conditions, post-conditions and loop invariants can be seen as
rich type annotations on a simple while-program. These logical
annotations can always be erased, leaving behind an untyped
while-program.

Like in Hoare logic, \ert{} distinguishes logical assertions from the
type-theoretic structure of the programming language. This is in
contrast to the traditional Curry-Howard interpretation of logic in
dependent type theory, and more closely resembles the \(\ms{Prop}\)
type of Coq (which is a sort of purely logical assertions), or even
more closely the "logic-enriched type theories" of Aczel and Gambino \cite{gambino-aczel}
(which extends type theory with a new judgement of logical
propositions).

Most dependent type theories also feature a notion of judgemental
equality, in which types are considered equal modulo some equational
theory (usually containing \(\beta\) and sometimes
\(\eta\)-conversions). \ert{} is designed to work without a
judgemental equality, since it is a source of both metatheoretical
difficulty, and complicates the design of tooling. However, there
are some dependent type theories, such as Objective Type Theory
\cite{objtt} and Zombie \cite{zombie}, which implement reduction
propositionally as axioms, similarly to what we have done.

\subsection{Refinement Logics and Squashed Curry-Howard}

As is well-known, even type theories without a \ms{Prop} sort like Coq
or Lean's still have a logical reading -- the famous "propositions as
types" principle, where each type-theoretic connective (function
types, pairs, sums) corresponds to a logical connective (implication,
conjunction, disjunction).

This is unsuitable for our purposes. We plan to use \ert{} as the
basis of extending practical SMT-based refinement type systems with
explicit proofs, and SMT solvers are fundamentally based on classical
logic. So we want the semantics of the propositions in our refinement
types to be classical as well, which is not possible when propositions
and types are identified.

This also makes it difficult to use the various modal techniques
proposed to integrate proof irrelevance into type theory, such
as Awodey and Bauer's squash types \cite{squash-types}, or Sterling and Harper's
phase modalities \cite{phase-modality}. 

For example, with the Awodey-Bauer squash type $[A]$, logical
disjunctions and existentials are encoded as
\begin{equation}
  \begin{array}{l}
    P \vee Q   = [P + Q]\\
    \exists x:X. P\, = [\Sigma x:X. P] 
  \end{array}
\end{equation}
where the $[-]$ operator has a degenerate equality. As a result, if
the type theory is intuitionistic the logic of propositions must be as
well, which is contrary to our needs. 

However, there is a deeper problem with using squash types. The
semantics of the Awodey-Bauer squash type is such that if $P$ is a
proposition (i.e., all inhabitants are equal), then $P$ and $[P]$ are
isomorphic.  Since $P$ is a proposition if all its inhabitants are
equal, a contractible type like $\Sigma x:\nats.\, x = n$ is a
proposition, and hence there is a map
$[\Sigma x:\nats. x = n] \to \Sigma x:\nats. x = n$.
 
That is, it is possible to extract computational data from a squashed
type, and so erasure of propositions and squashed types is a much more
subtle problem than it may first seem. Kraus \textit{et al}'s paper \textit{Notions
of Anonymous Existence in Martin-Löf Type Theory} \cite{anon-existence} studies this and
similar issues in detail.
There is a similar obstacle when using the Sterling-Harper approach,
which could use a pair of modalities to control whether a term is
potentially in the specification or runtime phases. Once again,
re-using type-theoretic connectives as logical connectives forces the
identification of the refinement logic and the type theory.

\subsection{Erasure in Dependent Type Theory}

Coq \cite{coq} supports a notion of erasure, in which terms of
proposition type are systematically elided as a dependently-typed
program is extracted to a functional language like Ocaml or
Haskell. It also lets users declare certain function parameters as
useless, but since Coq does not distinguish logical variables from
computational ones in its type system, a well-typed Coq term may fail
to successfully extract. In contrast, \ert{} can always erase both
proofs and logical variables, and furthemore guarantees that all
erased terms are also well-typed because it is a refinement over a
pre-existing simple type theory. This is not the case in Coq, where
extracted terms may have to use the unsafe cast \(\ms{Obj.magic}\). 

Another critical feature of \ert{} is that it supports Hoare-style
logical variables. is that they are not program variables, and cannot
influence the runtime behaviour of a program – they only exist for
specification purposes. So in a type
\begin{equation}
 \ms{vlen} : \forall n:\nats. \ms{Vec}\;n \to \{x : \nats \mid x = n\}
\end{equation}
where $\forall n:\nats$ is an intersection-style quantifier, it is
only possible to compute the length of the list by actually traversing
the list. In plain Martin L\"{o}f type theory, the corresponding type has a degenerate
implementation:
\begin{equation}
  \begin{array}{l}
    \ms{vlen} : \Pi n:\nats. \ms{Vec}\;n \to \Sigma x : \nats. x = n \\
    \ms{vlen}\;n\;\_ \equiv (n, \ms{refl}) 
  \end{array}
\end{equation}
This kind of computational irrelevance is different from what is
sometimes called proof-irrelevance or definitional irrelevance, since
different values of $n$ are not equal.

Two approaches that have arisen to manage this kind of irrelevance are
the implicit forall quantifier of the implicit calculus of
constructions (ICC) \cite{icc}, and the usage annotations in Atkey and McBride's
quantitative type theory (QTT) \cite{quantitative-type-theory}. 

As we did, the ICC introduced an intersection type $\forall x:T. U$ to
support computationally irrelevant quantification. However, because
there was no separation of the refinement layer from the base layer,
the denotational semantics of the ICC is much more complicated -- the
Luo-style extended calculus of constructions (ECC) \cite{ecc} has a
simple set-theoretic model, but the only known denotational model of
the ICC \cite{miquel-model} is based on coherence spaces. In our view,
this is a significant increase in the complexity of the model.

In contrast, we are able to model intersection types with ordinary
set-theoretic intersections. The reason for this is that the
refinement type discipline ensures that we only ever make use of
\emph{structural} set-theoretic operations. That is, every \ert{} type
is a subset of an underlying base type, and so when we take an
intersection, we are only taking the intersection of a family of
subsets of a particular set (the base type). From a mathematical point
of view, this is much better-behaved than taking intersections of
\emph{arbitrary} sets, and having this invariant lets us interpret
intersections more simply than is possible in the semantics of the
ICC.

Our semantics (and indeed, syntax) for intersection types is very
similar to the semantics of the "dependent intersection types"
introduced by \citet{dep-intersection} for the Nuprl system. They
worked with partial equivalence relations over a term model, rather
than our simple set-theoretic model. As mentioned above, we will
also need to move to a PER model to support function extensionality,
though we expect we can still consider PERs over sets, rather than
having to use a term model. 

In his PhD dissertation \cite{qtt-thesis}, Tejiščák studies how to
apply QTT directly to the problem of managing computational
irrelevance. The runtime and erasable annotations in QTT look very similar
to our distinction between computational and ghost variables. However,
we cannot directly use this type system, because it does not have
a sort of propositions, and therefore must express logical properties
in a construtive, Curry-Howard logic.

\subsection{Relation to Other Systems}

ATS~\cite{ATS} is a system in the Dependent ML style. That is, the type
system has a very hard separation between indices (which can
occur inside types) and terms (which do runtime computation).
Like $\ert$, it is possible in ATS to give 
explicit proofs of quantified and inductive formulae. 

ATS's erasure theorem is proved operationally, by exhibiting a
simulation between the reductions of the fully-typed language and
erased, untyped programs. This makes it hard to reason about the
equality of program terms (something like a logical relation would
have to be added), but ATS does not have to, because it distinguishes
program terms and indices, and only permits indices to occur in
types. 

However, since program terms cannot occur in constraints, correctness
arguments about functions are forced into an indirect style -- for
each recursive function, one must define an inductive relation
encoding its graph, and then show that inputs and outputs are related
according to this relation. As a result, proving something like (e.g.)
that the type of endofunctions $A \to A$, the identity, and function
composition have the structure of a monoid will be very
challenging. In $\ert$, in contrast, this would be
very straightforward, since (following liquid types) indices and
program terms are one and the same.

Modern ATS has also extended the proposition language to support
a notion of linear assertion, which permits verifying imperative
programs in the style of separation logic. Extending
$\ert$ with support for richer reasoning about
effectful programs is ongoing work.

\newcommand{\fstar}{\ensuremath{\mathrm{F}^{\ast}}}

\fstar is a full dependent type theory which has replaced the usual
conversion relation with an SMT-based approach. \fstar makes no effort
to keep quantifiers out of the constraints sent to its SMT solver, and
hence does not (and cannot) have any decidability guarantees -- the
\fstar typechecking problem is undecidable. This best-effort view lets
\fstar take maximum advantage of the solver, at the price of sometimes
letting the typechecker loop.  In contrast, $\ert$ has decidable, near
linear-time typechecking, because typing is fully syntax-directed and
has no conversion relation.

The treatment of computational irrelevance in \fstar is similar in
effect (though different in technical detail) to ICC. As in ICC,
ghost arguments can affect typing but not computations, but there
is no notion of a less-typed base language that ghosts can be
erased to. 

\bibliographystyle{ACM-Reference-Format}
\bibliography{references}

\clearpage 

\appendix

\section{Appendix}

\begin{figure}[H]
    \centering
    \begin{subfigure}{.49\textwidth}
        \begin{grammar}
            <A, B, C> ::= 
                \(\mb{1}\) \alt  
                (<x> ":" <A>) "\(\to\)" <B> \alt
                (<x> ":" <A>) "\(\times\)" <B> \alt
                <A> "+" <B> \alt
                \(\lrasm{u}{\varphi}{A}\) \alt
                \(\lrset{x}{A}{\varphi}\) \alt
                "\(\forall\)" <x> ":" <A>, <B> \alt
                "\(\exists\)" <x> ":" <A>, <B> \alt
                \(\nats\)
        \end{grammar}
        \caption{\ert{} types}
        \Description{
            \ert{} types
        }
        \label{fig:ert-types}
    \end{subfigure}
    \begin{subfigure}{.49\textwidth}
        \begin{grammar}
            <\(\varphi\), \(\psi\), \(\phi\)> ::=
                \(\top\) \alt
                \(\bot\) \alt
                \(\lrimp{u}{\varphi}{\psi}\) \alt
                (<u> ":" <\(\varphi\)>) "\(\land\)" <\(\psi\)> \alt
                <\(\varphi\)> "\(\lor\)" <\(\psi\)> \alt
                "\(\forall\)" <u> ":" <A>, <\(\varphi\)> \alt
                "\(\exists\)" <u> ":" <A>, <\(\varphi\)> \alt
                <a> \(=_{A}\) <b>
        \end{grammar}
        \vspace{4mm}
        \caption{\ert{} propositions}
        \Description{
            \ert{} propositions
        }
        \label{fig:ert-propositions}
    \end{subfigure}
    \begin{subfigure}{.49\textwidth}
        \begin{grammar}
            <a, b, c, e, e'> ::=
                \(()\) \alt
                <x, y, z> \alt
                \(\lambda\) <x> ":" <A>, <e> \alt
                <a> <b> \alt
                \((a, b)\) \alt
                \(\ms{let}\) \((x, y)\) ":" <A>
                "=" <e> \(\ms{in}\) <e'> \alt  
                \(\ms{inl}\) <e> \alt
                \(\ms{inr}\) <e> \alt
                \(\ms{cases}\) [<x> \(\mapsto\) <C>] <e>
                    (\(\ms{inl}\) <y> "\(\mapsto\)" <a>)
                    (\(\ms{inr}\) <z> "\(\mapsto\)" <b>) \alt
                \(\lambda\) <x> ":" <\(\varphi\)>, <e> \alt
                <a> <p> \alt
                \(\{a, p\}\) \alt
                \(\ms{let}\)
                \(\{x, u\}\) ":" <A>
                "=" <e> \(\ms{in}\) <e'> \alt 
                \(\lambda\) \(\|x : A\|\), <e> \alt
                <a> \(\|b\|\) \alt
                (\(\|\) <a> \(\|\), <b>) \alt
                \(\ms{let}\)
                \((\|x\| , y)\) ":" <A> 
                "=" <e> \(\ms{in}\) <e'> \alt
                \(0\) \alt
                \(\ms{succ}\) \alt
                \(\ms{natrec}\) [<x> \(\mapsto\) <C>] <e> <a> 
                (\(\|\)\(\ms{succ}\) <y>\(\|\), <z> \(\mapsto\) <b>) \alt
                %\(\ms{subst}\)[<x> \(\mapsto\) <B>][<a>][<b>] <p> <e> \alt
                \(\ms{absurd}\) <p>
        \end{grammar}
        \caption{\ert{} terms}
        \Description{
            \ert{} terms
        }
        \label{fig:ert-terms}
    \end{subfigure}
    \begin{subfigure}{.49\textwidth}
        \begin{grammar}
        <p, q, r> ::=
            \(\langle\rangle\) \alt
            <u, v, w> \alt
            \(\ms{absurd}\) <p> \alt
            \(\hat{\lambda} x : \varphi\), <p> \alt
            <p> <q> \alt
            \(\langle p, q \rangle\) \alt
            \(\ms{let}\) \(\langle x, y \rangle\) ":" <\(\varphi\)>
            "=" <p> \(\ms{in}\) <q> \alt 
            \(\ms{orl}\) <p> \alt
            \(\ms{orr}\) <q> \alt
            \(\ms{cases}_{\ms{or}}\) 
                [<u> \(\mapsto\) <\(\phi\)>]
                <p>
                (\(\ms{orl}\) <y> \(\mapsto\) <q>)
                (\(\ms{orr}\) <z> \(\mapsto\) <r>) \alt
            \(\hat{\lambda}\) \(\|\) <x> ":" <A> \(\|\), <p> \alt
            <p> \(\|a\|\) \alt
            \(\langle\|a\|, p\rangle\) \alt
            \(\ms{let}\) \(\langle\|x\|, u\rangle\) ":" <\(\varphi\)>
            "=" <p> \(\ms{in}\) <q> \alt
            \(\ms{let}\) \((x, y)\) ":" <A>
            "=" <p> \(\ms{in}\) <q> \alt
            \(\ms{let}\)
            \(\{x, y\}\) ":" <A>
            "=" <p> \(\ms{in}\) <q> \alt
            \(\ms{let}\)
            \((\|x\| , u)\) ":" <A> 
            "=" <p> \(\ms{in}\) <q> \alt
            \(\ms{subst}\)[<x> \(\mapsto\) <B>][<a> \(\to\) <b>] <p> <q> \alt
            \(\ms{cases}\)
            [<x> \(\mapsto\) <\(\varphi\)>]
            (\(\ms{inl}\) <y> "\(\mapsto\)" <p>)
            (\(\ms{inr}\) <z> "\(\mapsto\)" <q>) \alt
            \(\ms{ind}\) [<x> \(\mapsto\) <\(\varphi\)>] <e> <p> 
            (\(\ms{succ}\) <y>, <u> \(\mapsto\) <q>) \alt
            <axiom>
        \end{grammar}
        \caption{\ert{} proofs}
        \Description{
            \ert{} proofs
        }
        \label{fig:ert-proofs}
    \end{subfigure}
    \begin{subfigure}{.49\textwidth}
        \begin{grammar}
            <axiom> ::=
                \(\ms{rfl}\) <a> \alt
                \(\ms{discr}\) <a> <b> <p> \alt
                \(\ms{uniq}\) <a> \alt
                <beta> \alt
                \(\eta_{\ms{ty}}\) <e> \alt
                \(\ms{ext}_{\ms{ty}}\) <f> <g> <i> <p> \alt
                \(\ms{ext}_{\ms{pr}}\) <f> <g> <i> <p> \alt
                \(\ms{ir}_{\ms{pr}}\) [<x> \(\mapsto\) <e>] <p> <q> \alt
                \(\ms{ir}_{\ms{ty}}\) <e> <a> <b>
        \end{grammar}
        \vspace{16.8mm}
        \caption{\ert{} axioms}
    \end{subfigure}
    \begin{subfigure}{.49\textwidth}
        \begin{grammar}
            <beta> ::=
                \(\beta_{\ms{pr}}\) (<x> \(\mapsto\) <e>) <p> \alt
                \(\beta_{\ms{ty}}\) (<x> \(\mapsto\) <e>) <a> \alt
                \(\beta_{\ms{ir}}\) (<x> \(\mapsto\) <e>) <a> \alt
                \(\beta_{\ms{left}}\) 
                    [<x> \(\mapsto\) <C>]
                    (\(\ms{inl}\) <y> "\(\mapsto\)" <a>)
                    (\(\ms{inr}\) <z> "\(\mapsto\)" <b>)
                    (\(\ms{inl}\;c\)) \alt
                \(\beta_{\ms{right}}\) 
                    [<x> \(\mapsto\) <C>]
                    (\(\ms{inl}\) <y> "\(\mapsto\)" <a>)
                    (\(\ms{inr}\) <z> "\(\mapsto\)" <b>)
                    (\(\ms{inr}\;c\)) \alt
                \(\beta_{\ms{zero}}\) [<x> \(\mapsto\) <C>] <a> 
                (\(\|\)\(\ms{succ}\) <y>\(\|\), <z> \(\mapsto\) <b>) \alt
                \(\beta_{\ms{succ}}\) [<x> \(\mapsto\) <C>] (\(\ms{succ}\) <e>) <a>  
                (\(\|\)\(\ms{succ}\) <y>\(\|\), <z> \(\mapsto\) <b>) \alt
                \(\beta_{\ms{pair}}\) (<a>, <b>) ((<y>, <z>) \(\mapsto\) <e>) \alt
                \(\beta_{\ms{set}}\) \{<a>, <p>\} (\{<y>, <z>\} \(\mapsto\) <e>) \alt
                \(\beta_{\ms{repr}}\) \((\|a\|, b)\) ((<y>, <z>) \(\mapsto\) <e>)
        \end{grammar}
        \caption{\ert{} \(\beta\)-rules}
    \end{subfigure}
    \caption{Grammar for \ert{}.}
    \Description{
        \ert{} grammar
    }
    \label{fig:ert-grammar}
\end{figure}

\begin{figure}
    \centering
    \begin{minipage}{.5\textwidth}
    \centering
    \begin{grammar}
        <A, B, C> ::= 
            \(\mb{0}\) | \(\mb{1}\) \alt  
            <A> "\(\to\)" <B> \alt
            <A> "\(\times\)" <B> \alt
            <A> "+" <B> \alt
            \(\nats\)
    \end{grammar}
    \vspace{33mm}
    \end{minipage}%
    \begin{minipage}{.5\textwidth}
    \centering
    \begin{grammar}
        <a, b, e, l, r> ::=
            () \alt
            \(\ms{error}\) \alt
            \(\lambda\) <x>: <A>. <e> \alt
            <a> <b> \alt
            (<l>, <r>) \alt
            \(\ms{let}\) (<x>, <y>) = <e> \(\ms{in}\) <a> \alt
            \(\ms{inl}\) <a> \alt
            \(\ms{inr}\) <a> \alt
            \(\ms{cases}\) <e>
            (\(\ms{inl}\) <x> \(\mapsto\) <l>)
            (\(\ms{inl}\) <y> \(\mapsto\) <r>) \alt
            0 \alt
            \(\ms{succ}\) \alt
            \(\ms{natrec}\) <e> <a> (<x> \(\mapsto\) <b>) \alt
            \(\ms{let}\) <x> = <a> \(\ms{in}\) <b>
    \end{grammar}
    \end{minipage}
    \caption{Grammar for the \stlc{} with \(\ms{error}\)}
    \Description{
        A grammar for the STLC with \(\ms{error}\)
    }
    \label{fig:stlc-grammar}
\end{figure}

\end{document}